\documentclass[reprint]{revtex4-1}

\usepackage{amsmath, amssymb, amsfonts}
\usepackage{graphicx}
\usepackage{color}
\usepackage{float}
\usepackage{wrapfig}
\usepackage{array}
\usepackage{hyperref}

\hypersetup{
    colorlinks = false,
   urlcolor={1 1 1}
    }

\begin{document}
\title{Evolution of network architecture in a granular material under compression}
\author{Lia Papadopoulos}
\affiliation{Department of Physics \\ University of Pennsylvania}
\author{James G. Puckett}
\affiliation{Department of Physics\\Gettysburg College}
\author{Karen E. Daniels}
\affiliation{Department of Physics\\North Carolina State University}
\author{Danielle S. Bassett}
\affiliation{Departments of Bioengineering and Electrical \& Systems Engineering \\University of Pennsylvania}
\email{dsb@seas.upenn.edu}
\date{\today}

\newpage

\begin{abstract}

As a granular material is compressed, the particles and forces within the system arrange to form complex and heterogeneous collective structures. Force chains are a prime example of such structures, and are thought to constrain bulk properties such as mechanical stability and acoustic transmission. However, capturing and characterizing the dynamic nature of the intrinsic inhomogeneity and mesoscale architecture of granular systems can be challenging. A growing body of work has shown that graph theoretic approaches may provide a useful foundation for tackling these problems. Here, we extend the current approaches by utilizing \textit{multilayer networks} as a framework for directly quantifying the evolution of mesoscale architecture in a compressed granular system. We examine a quasi-two-dimensional aggregate of photoelastic disks, subject to biaxial compressions through a series of small, quasistatic steps. Treating particles as network nodes and inter-particle forces as network edges, we construct a multilayer network for the system by linking together the series of static force networks that exist at each strain step. We then extract the inherent mesoscale structure from the system by using a generalization of \textit{community detection} methods to multilayer networks, and we define quantitative measures to characterize the reconfiguration and evolution of this structure throughout the compression process. We separately consider the network of normal and tangential forces, and find that they display different structural evolution throughout the compresion process. To test the sensitivity of the network model to particle properties, we examine whether the method can distinguish a subsystem of low-friction particles within a bath of higher-friction particles. We find that this can be done by considering the network of tangential forces. The results discussed throughout this study suggest that these novel network science techniques may provide a direct way to compare and classify data from systems under different external conditions or with different physical makeup.
\end{abstract}

\maketitle

\section{Introduction}

Granular materials \cite{Heinrich:1996} come in many forms, from soils, sands, and grains to powders and pharmaceuticals. However, despite their prevalence, there are still open questions about how the seemingly simple interactions of contact forces lead to the observed emergent behavior of these systems. An active area of research lies in understanding the mechanisms that govern deformation in granular materials subjected to compression and shear. Under both of these external perturbations, the force network of granular systems exhibits complex and inhomogeneous structure in the form of strongly interacting collections of particles known as \textit{force chains} (see Fig.~\ref{f:experiment}\emph{C}) \cite{Dantu:1957, Liu:1995aa, Radjai:1996aa, Mueth:1998aa, Howell:1999aa, Majmudar:2005aa,Peters:2005aa}. This architecture is thought to constrain the mechanical properties and stability of granular materials \cite{Howell:1999aa,Radjai:1998aa,Radjai:1999a,Majmudar:2005aa} and may also be responsible for nonlinear and heterogeneous features of acoustic signal transmission \cite{Hidalgo:2002a, Makse:1999aa, Owens:2011a, Goddard:1990a, Liu:1992aa, Liu:1993aa, Liu:1994}.

Two notable features of force chains are that they are {\it mesoscale} structures, intermediately sized between the particle scale and the system scale, and that their physical structure depends on the loading history \cite{Majmudar:2005aa}. These characteristics present a challenge, as there is currently no single modeling framework that explicitly addresses the presence of mesoscale architecture in particulate systems and how it reconfigures under external influences. The development of such models is critical, as particulate and continuum methods cannot fully describe the observed properties exhibited by these systems \cite{Makse:1999aa}.

Recently, a number of studies have suggested that graph theoretic \cite{Albert:2002aa,Newman:2003aa, Newman:2010aa} approaches provide a powerful and natural paradigm in which to study granular media. Many of these analyses have focused on the characterization of discrete sets of static granular force networks throughout compression \cite{Arevalo:2009aa, Arevalo:2010aa, Arevalo:2010ba, Tordesillas:2010aa, Walker:2010aa, Bassett:2012aa}, tapping \cite{Arevalo:2013aa}, or tilting \cite{Smart:2008aa}, using traditional graph metrics such as degree, clustering coefficients, and cycles of different lengths. Other work has probed the dynamical nature of sheared systems by considering time-evolving networks of broken-links \cite{Herrera:2011aa, Slotterback:2012aa}, and grain property networks have been used to understand rearrangements in discrete element simulations of compressed systems \cite{Walker:2012aa}. Methods from algebraic topology, and in particular, persistent homology \cite{Ghrist:2008aa,Carlsson:2009aa} have also been used to quantify the evolution of force networks, providing important insights into the nature of compressed \cite{Kondic:2012aa, Kramar:2013aa,Kramar:2014aa} and tapped \cite{Ardanza-Trevijano:2014aa, Pugnaloni:2015, Kondic:2015} granular materials. 

One reason for the promise of network-based measures is that they can assess material architecture at a range of length scales, including the important mesoscale regime. Recent work by \textcite{Bassett:2015aa} showed that a network clustering technique known as \textit{community detection} could be used to extract the underlying mesoscale force chain structure from static granular networks. In this study, we extend the static model, and suggest that \textit{multilayer networks} may be a particularly promising framework in which to simultaneously examine the mesoscale architecture of granular systems, as well as to probe network evolution and reconfiguration in a straightfoward manner. This novel approach has thus far been unexplored. 

Multilayer networks encompass several different types of complex graph constructions, and the word can take on a number of meanings depending on the context (see \cite{Kivela:2014aa, Boccaletti:2014aa} for comprehensive reviews). For example, a multilayer network may capture different types of connections between nodes, may quantify interactions between different systems, or may be used to study dynamical processes that occur across time. Here, we focus on a specific subset of these possibilities. In particular, we restrict ourselves to \textit{temporal networks} with \textit{diagonal} and \textit{ordinal} inter-layer couplings. A \textit{temporal} network consists of a sequential series of static graphs (the layers) ordered such that time-dependence is accounted for. \textit{Diagonal} couplings mean that a node in one layer is only connected to itself in other layers. Finally, \textit{ordinal} inter-layer couplings only allow connections between layers that are adjacent to each other in time. In this study, we are interested in the evolution of a granular material as it undergoes biaxial compression. We thus represent a snapshot of the system at a particular point in its evolution as a spatially embedded graph where particles are nodes and inter-particle forces are weighted edges. Repeating this process at several discrete strain steps yields an ordered set of static networks, which can then be combined into a single multilayer graph with the temporal dependence set by the order of the strain steps. 

We develop and apply this multilayer network formalism to experimental granular data \cite{Puckett:2013aa} to (i) extract evolving mesoscale structure from the force network, (ii) understand how this architecture reconfigures throughout compressive (strain) steps, (iii) uncover physical properties of evolving mesostructures and relate them to measures of network rearrangement, and (iv) examine the impact of inter-particle friction on multilayer mesostructures. To achieve this, we represent an ensemble of granular packings as multilayer graphs using the set of force networks (both normal and tangential) obtained from each step above the jamming point. We then use \textit{multilayer community detection} to extract groups of particles that evolve together throughout the compression procedure. This method allows us to directly characterize the progression of inherent mesoscale organization as a function of strain step.

The outline of this paper is as follows. In Sec.\ref{s:experimental_methods} we describe the granular experiments. Sec.~\ref{s:math_model} is dedicated to an explanation of the multilayer network model and community detection, which lay the theoretical foundations for the rest of the paper. A series of results describing the granular network evolution are presented in Sec.~\ref{s:results}, and in Sec.~\ref{s:discussion} we discuss broader implications of our method and findings, and directions for future work.

\section{Experimental Methods \label{s:experimental_methods}}

\begin{figure*}[floatfix]
\centering
\includegraphics[width=0.8\linewidth]{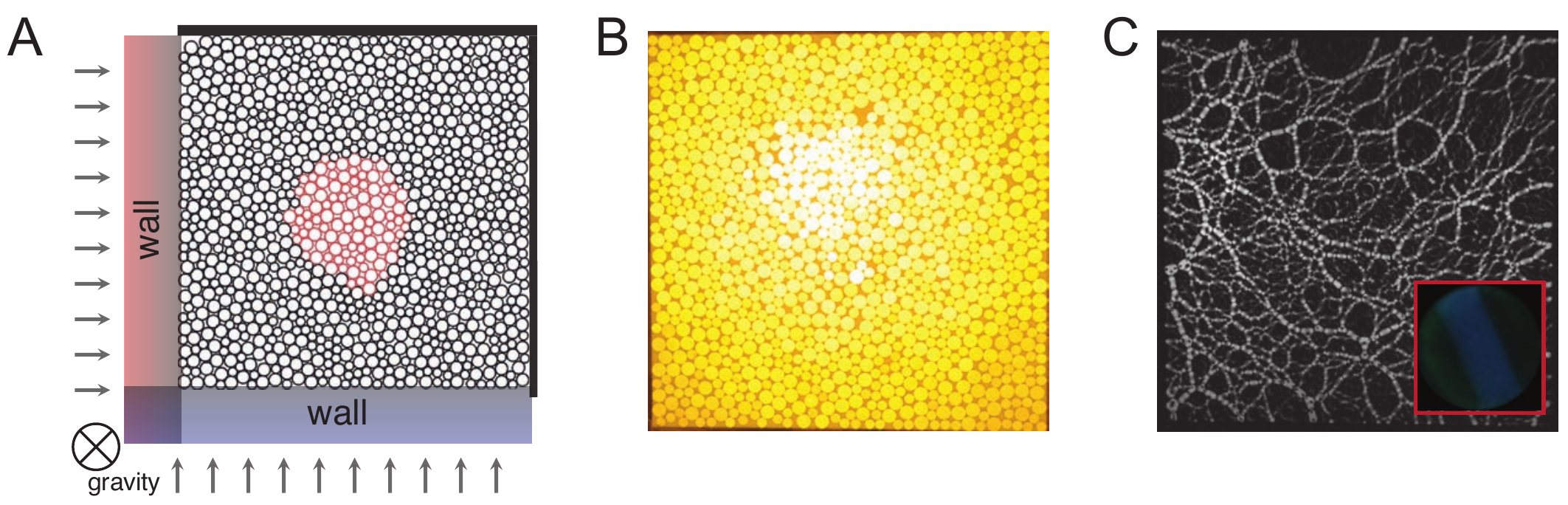}
\caption{\label{f:experiment} \textbf{Experimental setup}. \emph{(A)} A schematic of the apparatus showing two walls bi-axially compressing an array of disk-shaped particles composed of an outer subsystem (black, high $\mu$) and an inner subsystem (red, low $\mu$). \emph{(B)} The first image is taken with unpolarized white light and is used to locate particle positions. \emph{(C)} The network of the photoelastic stress pattern, visualized with a second image taken with polarized light. This image allows for the calculation of the normal and tangential forces at each contact. A third image (inset) taken with ultraviolet light is used to identify the subsystem particles, which are tagged with fluorescent ink. }
\end{figure*}

We study the biaxial compression of a granular monolayer on a nearly frictionless surface provided by an air table (Fig.~\ref{f:experiment}). The system is composed of an inner subsystem and an outer bath which differ only in the inter-particle friction coefficient.  At the beginning of each compression cycle, the system is in a dilute, unjammed state and is then compressed quasistatically until it is highly stressed.  By repeating this protocol many times, we generate an ensemble of configurations for which we record particle positions (Fig.~\ref{f:experiment}\emph{B}), and use photoelastic measurements (Fig.~\ref{f:experiment}\emph{C}) to calculate contact forces \cite{Puckett}. The presence of the subsystem and bath allows us to characterize both the system as a whole, and to investigate the physical differences that exist between the high- and low-friction regions. 

The inner subsystem and bath are composed of 100 and 904 particles, respectively.  The two systems are both composed of a bidisperse mixture of disks (diameters $d_{s}$ = 11 mm and $d_{l}$  =  15.4 mm, in equal concentrations), which differ only in the inter-particle friction coefficient. The friction coefficient for the inner subsystem is $\mu < 0.1$ and for the high friction bath is $\mu \approx 0.8$. 

A single CCD camera located above the apparatus records three separate images of the system: unpolarized white light for the particle positions, polarized light for recording the photoelastic response, and ultraviolet light for identifying the subsystem particles, as shown in Fig.~\ref{f:experiment}.  Using the photoelastic image, we measure the normal and tangential forces at each contact on each disk in the assembly.  To calculate the forces, we use a nonlinear least squares optimization algorithm \cite{Levenberg1944,Marquardt1963} to minimize the error between the observed and fitted image of the particle. Details and source code are available for download at \cite{Puckett}. The third image, taken using a black-light illumination, identifies which particles are low-friction via their fluorescent marking. 

Each compression cycle begins with an initial dilute state with global packing fraction $\Phi \sim 0.6$.  In the dilute state, the granular system is unjammed and unable to support stress.  Two walls then biaxially compress the system in a series of small steps of constant size ($\Delta\Phi  = 0.009$, equivalently $\Delta x  = 0.3$~mm or 0.02$r_{l}$). As the system is compressed via these discrete, quasistatic steps, we observe the percolation of force chains throughout both the bath and the subsystem at a value $\Phi_\mathrm{J}$. This is the onset of rigidity, and as the system is further compressed beyond this point, the contact forces grow in strength and the average number of contacts per particle increases.  For each of the 97 configurations we locate $\Phi_\mathrm{J}$ as the step at which photoelastic signals are first present, and consider the evolution of the system due to the subsequent applied strain steps. Further details of the experimental apparatus and measurements were published previously \cite{Puckett:2013aa}.  

\subsection{Particle tracking}

It is important to note that the construction of a multilayer network requires knowledge about which node (particle) is which from one layer (strain step) to the next \cite{Kivela:2014aa,Boccaletti:2014aa}. Under the dynamics described above, we are assured that no particles are removed from or added to the system, and we similarly require that the multilayer network also has this constraint. (In general, multilayer graphs can indeed be constructed for growing systems, where the number of nodes is constantly changing.) In order to correctly identify the particles in each layer, we use the Blair-Dufresne particle tracking algorithm \cite{particle_tracking}. This algorithm, implemented in MATLAB,  requires the choice of a ``displacement'' parameter, which is an estimate of the maximum distance that a particle moves between consecutive frames.  As we do not expect large particle displacements in the jammed packings, we initialize the tracking with a displacement parameter value that is much less than the minimum particle diameter $d_{s}$, and increase the allowed displacement in small steps until all particles are tracked consistently across all steps.

\section{Mathematical Model \label{s:math_model}}

\subsection{Multilayer network representation \label{s:multi_network}}

\begin{figure*}[floatfix]
\centering
\includegraphics[width=\textwidth]{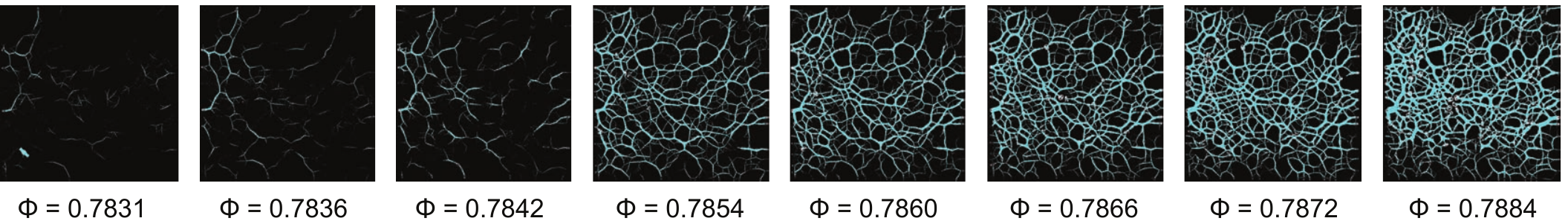}
\caption{\label{f:photoelastic_evolution} \textbf{Evolution of the force network}. Photoelastic images of the system, taken at a sequence of eight steps above $\Phi_{J} \approx 0.7825$.  The corresponding network of (normal) forces is overlaid on top of each picture. The collection of these static networks may together be represented as a single multilayer network that captures the evolution of the granular system throughout compression.}
\end{figure*}

In general, \textit{temporal} multilayer networks are mathematical objects that describe and quantify the evolution of networked systems as a function of time or some other variable of interest \cite{Kivela:2014aa,Boccaletti:2014aa}. The construction of such a multilayer network first requires a set of individual networks that describe the system at discrete steps in its evolution. These static ``layers'' can each be represented as an adjacency matrix, $\mathbf{A}$, describing the connectivity and weight connecting the nodes in the given layer. The set of adjacency matrices may then be combined into a rank-3 adjacency tensor, $\mathcal{A}$ \cite{Domenico:2013aa}, to form a multilayer network representation of the system. The elements of $\mathcal{A}$ are defined such that 
\begin{equation}
\begin{split}
\mathcal{A}_{ijl} = \left\{ \begin{array}{ll}
        a_{ijl}  & \mbox{if node $i$ and $j$ are connected} \\
        & \mbox{in layer $l$} \\
        0 &  \mbox{otherwise},\end{array} \right. \
 \end{split}
\label{eq:adjacency_tensor_general}
\end{equation}
where $a_{ijl}$ is the weight between nodes $i$ and $j$ in layer $l$, and $L$ is the total number of layers (steps). 

For the present situation, we are interested in how the force network of a granular packing reconfigures through a sequence of compressive steps above $\Phi_\mathrm{J}$. We thus wish to represent the system as an ordered, temporal multilayer network, which captures the evolution of the system as a function of compression. To form such a network, we thus let nodes be particles, weighted edges be the forces between contacting particles, and the number of strain steps be the third dimension across which we observe changes in network structure. Using the force information obtained from the photoelastic disk experiments, we construct two multilayer force networks for a given experimental run, one using the normal forces, $F^{n}$, between particles and another using the magnitude of the tangential forces $|F^{t}|$. Specifically, we denote the normal force adjacency tensor as $\mathcal{A}_{ijl}^{n}$, defined as in Eq.~\ref{eq:adjacency_tensor_general} with $a_{ijl} = f^{n}_{ijl}$, where $f^{n}_{ijl}$ is the normal force between particles $i$ and $j$ at step $l$. Similarly, we write the tangential force adjacency tensor as $\mathcal{A}_{ijl}^{t}$, defined as in Eq.~\ref{eq:adjacency_tensor_general} with $a_{ijl} =|f^{t}_{ijl}|$, where $f^{t}_{ijl}$ is the tangential force between particles $i$ and $j$ at step $l$. This process is repeated for all of the experimental configurations, resulting in a large ensemble of multilayer networks.  In Fig.~\ref{f:photoelastic_evolution}, we show an example set of photoelastic images with the corresponding network of normal forces overlaid.

In addition to the \textit{intra}-layer weights, multilayer networks have a second set of \textit{inter}-layer edges that connect nodes in different layers of the network. In this study, we consider only \textit{diagonal} and \textit{ordinal} inter-layer couplings, meaning that a given particle is connected only to itself (i.e. diagonal coupling) inter-layer edges exist only between \textit{adjacent} layers (i.e ordinal coupling). The inter-layer weights are crucial to the structure of the network; in Sec.~\ref{s:choosing_omega}, we discuss how these couplings are chosen for the granular system at hand.

This graphical construction is a powerful approach. Importantly, each layer of the adjacency tensor encodes both the topology (connectivity) as well as the strength of interactions between particles in the system at a given packing fraction. We will see that the extension to a \textit{multilayer} framework not only allows one to study the static behavior of granular packings, but also promotes a direct investigation and characterization of the evolution of the system throughout the compressive steps.

\subsection{Community detection}

A compelling reason to use graphical representations of spatially embedded systems \cite{Barthelemy:2011} is that network theoretic approaches provide a means to extract and characterize organization that is present at a range of length scales. Granular materials are a prime example of a complex system in which multiple length scales are relevant to a full understanding of the system and to a prediction of bulk properties. For example, the fundamental interactions in these systems are local in the sense that they occur between nearest neighbor particles. But under compression, those local interactions lead to important heterogeneous structure at the \textit{mesoscale} in the form of force chains \cite{Dantu:1957, Liu:1995aa, Radjai:1996aa, Mueth:1998aa, Howell:1999aa, Majmudar:2005aa,Peters:2005aa, Ostojic:2006aa}. 

In the present study, we aim to both extract mesoscale structures from the force network, and then characterize the reconfiguration that occurs in these structures due to the applied strain steps. In order to accomplish this, we build upon recent approaches that utilize \textit{community detection} techniques to extract groups of strongly interacting particles from the granular network \cite{Bassett:2012aa,Bassett:2015aa}, generalizing the methods to the multilayer regime.

A \textit{community} or \textit{module} in a network is a set of nodes that are densely interconnected amongst themselves, and relatively weakly connected to other nodes \cite{Porter:2009aa}. The extraction of community structure is of general interest in network science, as it is thought that these mesoscale units are important to the function of many real systems \cite{Porter:2009aa}. Several community detection methods exist \cite{Porter:2009aa, Fortunato:2010aa}; here, we apply the popular method of \textit{modularity maximization}, whereby nodes are partitioned into communities via maximization of a quality function known as \textit{modularity} \cite{Newman:2004ab,Newman:2006ab}.

\subsubsection{Single layer modularity maximization}

\begin{figure*}
\includegraphics[width=\textwidth]{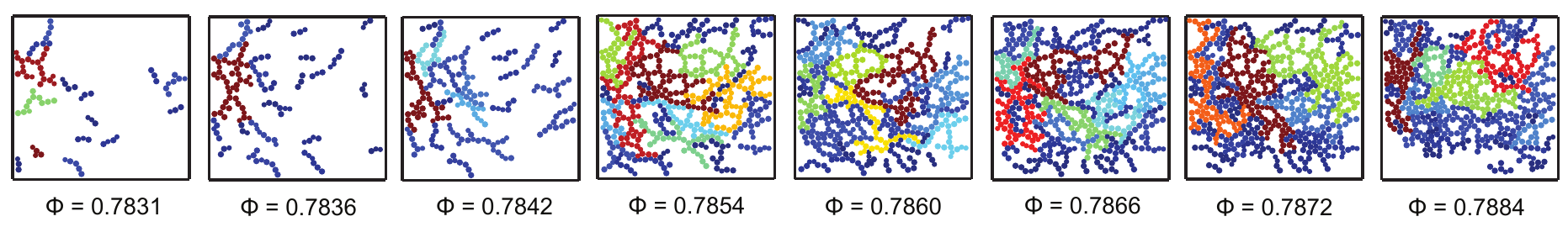}
\centering
\caption{\label{f:static_norm_comms} \textbf{An example of single layer community structure.} When single layer community detection is performed on the series of normal force networks above $\Phi_{J}$, the groups of particles at each step are highly reminiscent of force chains. However, there is no notion of linking these mesostructures from one compressive step to the next.}
\end{figure*}

In a single layer network with adjacency matrix \textbf{A}, modularity is given by
\begin{equation}
Q_{sing} = \frac{1}{2m} \sum_{ij} [A_{ij} - \gamma P_{ij}] \delta(c_{i},c_{j}),
\label{eq: static_modularity}
\end{equation}
where $c_{i}$ is the community of node $i$, $c_{j}$ is the community of node $j$, $P_{ij}$ is the expected edge weight between nodes $i$ and $j$ under a specified null model, and $\gamma$ is the \textit{structural resolution parameter}. In the overall normalization, $ m = \frac{1}{2} \sum_{ij} A_{ij} $, which is the weighted degree or \textit{strength} of the network. The structural resolution parameter $\gamma$ allows for the control of size and number of communities: smaller $\gamma$ leads to larger communities, and larger $\gamma$ leads to smaller communities. Maximization of $Q_{sing}$ with respect to the assignment of nodes to communities yields a partition in which intra-community connections are as strong as possible relative to the null model. It is important to note, however, that modularity maximization is NP-hard \cite{Good:2010aa} and should therefore be repeated several times for the same network and set of parameters in order to obtain an ensemble of optimizations \cite{Bassett:2013aa}. Throughout this work, we use a Louvain-like locally greedy algorithm for modularity maximization \cite{Blondel:2008aa, Jutla:GenLouvain}.

\subsubsection{A physically-informed null model}

A proper choice of null model is vital to community detection techniques, as it affects both the interpretation and utility of the community structures obtained \cite{Bassett:2013aa,Sarzynska:2015aa}. The most commonly used null model in the literature is the Newman-Girvan (NG) model \cite{Newman:2004ab,Newman:2006ab}, which in the case of a static network is given by $P_{ij} = \frac{k_{i} k_{j}}{2m}$, where $ k_{i} = \sum_{j} A_{ij} $ is the strength of node $i$, and $m$ is the total strength of the network, as before. In the Newman-Girvan model (which is also sometimes referred to as the \emph{configuration model}), $P_{ij}$ gives the expected edge weight between nodes $i$ and $j$ in a randomized network that has the same weighted degree distribution as the real network. As a randomized version of the real graph, the NG model is most appropriate to use as a null model in situations where all connections between nodes are at least \textit{possible}. In many physical or spatially embedded systems, however, this is not the case, and there are constraints that prevent the existence of several edges. For example, in the granular networks considered here, edges can only exist between nearest neighbor particles, and it is imperative to consider this fact when designing a null model for these types of systems. A better choice in this instance is the physically informed \textit{geographical null model} \cite{Bassett:2015aa, Bassett:2013aa}, defined to be
\begin{equation}
P_{ij} = \langle f \rangle B_{ij},
\label{eq: geog_null}
\end{equation}
where $\langle f \rangle$ is the average inter-particle force (either normal or tangential) in the network and $B_{ij}$ is the contact matrix, with elements
\begin{equation}
B_{ij} = \left\{ \begin{array}{ll}
        1& \mbox{if particle $i$ and $j$ are in contact} \\
        0 & \mbox{otherwise},\end{array} \right . \
\label{eq:contact_matrix}
\end{equation}
Because this null model maintains the contact structure of the real network, it importantly takes into account the physical constraints on the possible patterns of connectivity between particles. Additionally, it selects for strongly connected sets of particles carrying forces larger than $\gamma \langle f \rangle$.

\subsubsection{Multilayer modularity maximization \label{s:multi_mod_max} } 

\begin{figure}[floatfix]
\includegraphics[width=\columnwidth]{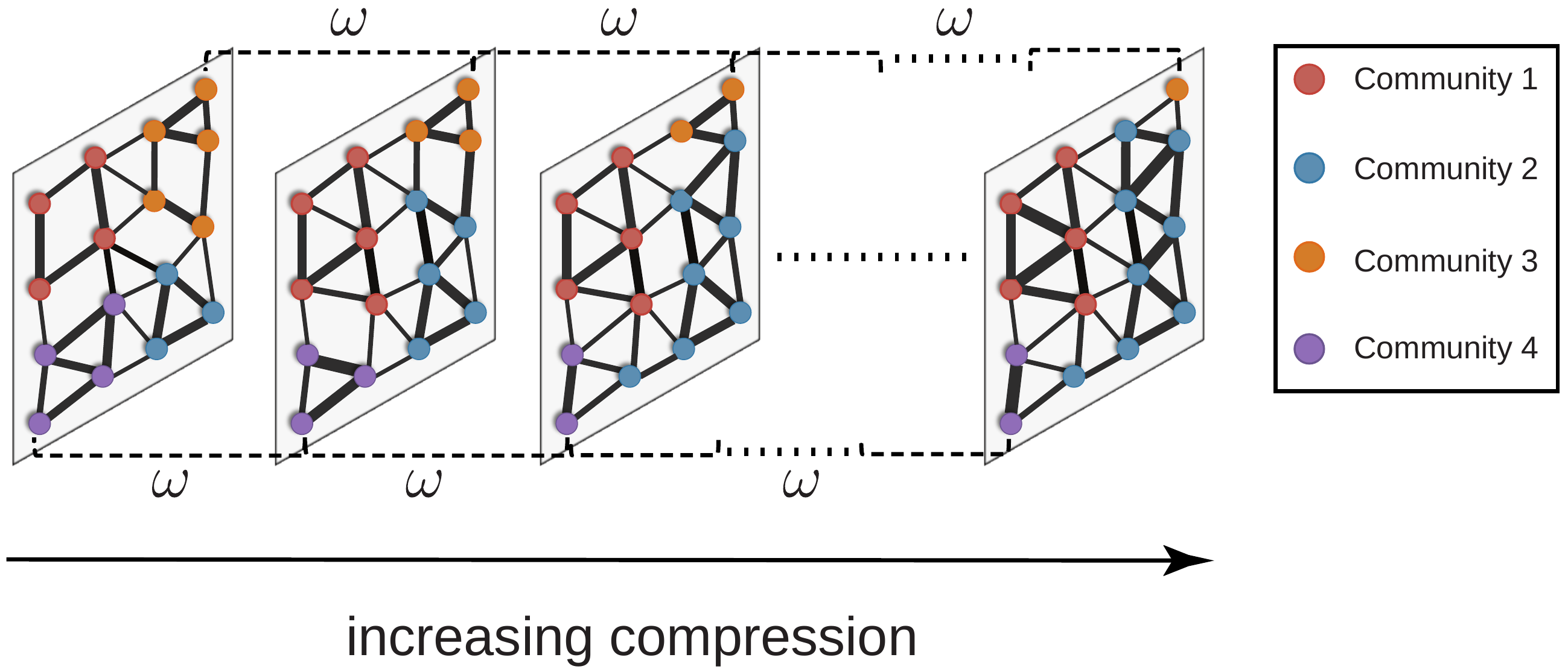}
\centering
\caption{\label{f:multilayer_schematic} \textbf{A schematic of a multilayer network with layer-dependent community structure.} Each layer represents a static granular force network in which nodes (particles) are connected to one another via intra-layer weighted edges. These weights can be either the normal contact forces or the absolute value of the tangential contact forces. Additionally, the same particle in consecutive layers is linked to itself with an inter-layer coupling, $\omega$. For clarity, we only show two such couplings, but these inter-layer edges exist between all particles (and across all layers). Evolving communities are extracted from the multilayer network to study the mesoscale organization in the system, and to understand how it reconfigures due to the compressive cycle. The community structure can be determined from the network by maximizing the multilayer modularity, $Q_{multi}$ (Eq.~\ref{eq:dynamic_modularity}). 
In this schematic, the particles belonging to different communities are labeled by different colors. Note that the same community can persist across all layers and reconfigure in terms of particle content and strength throughout compression.}
\end{figure}

When community detection is performed on a series of single layer granular networks, such as those obtained here after each applied strain step, the result is a set of \textit{independent} partitions of particles into communities at each step (Fig.~\ref{f:static_norm_comms}). As demonstrated in \cite{Bassett:2015aa}, the communities at a given step correspond to sets of strongly interacting groups of particles, and are clearly reminiscent of physical force chains observed in the photoelastic disk experiments. However, in this scheme, the community structure is not in any way linked from one layer to the next, barring any notion of continuation. Instead, the communities are treated as independent from one another, which is an inaccurate representation of the physics and further challenges our ability to directly capture the evolution of network structure and reconfiguration at the mesoscale.

To form a more dynamic picture of structural network evolution, we investigate the community structure of \textit{multilayer} granular force networks by applying the recent generalization of modularity maximization to temporal networks \cite{Mucha:2010aa, Bazzi:2016}. In this formulation, the multilayer modularity is defined to be
\begin{equation}
Q_{multi} = \frac{1}{2 \mu} \sum_{ijls} [(\mathcal{A}_{ijl} - \gamma_{l}\mathcal{P}_{ijl}) \delta_{lm} + \omega_{jlm}\delta_{ij}] \delta(c_{il},c_{jm})],
\label{eq:dynamic_modularity}
\end{equation}
where $\mathcal{A}_{ijl}$ is the $(i,j)$ component of the $l^{th}$ layer of the adjacency tensor, $\mathcal{P}_{ijl}$ is the $(i,j)$ component of the $l^{th}$ layer of the null model tensor, and $\gamma_{l}$ is the structural resolution parameter for layer $l$. In addition to $\gamma$, the multilayer modularity requires another free parameter, $\omega$ (often referred to as an \textit{inter-layer coupling} or \textit{temporal resolution parameter}) which sets the the strength of connections between layers. Namely, $\omega_{jlm}$ is the strength of the coupling that links node $j$ in layer $l$ to itself in an adjacent layer $m$ (i.e. the diagonal and ordinal coupling). The quantities $c_{il}$ and $c_{jm}$ are the community assignments of node $i$ in layer $l$ and node $j$ in layer $m$, respectively. Defining the intraslice strength of node $j$ in layer $l$ as $k_{jl} = \sum_{i} A_{ijl}$ and the strength of node $j$ across layers as  $w_{jl} = \sum_{m}{\omega_{jlm}}$, then the multilayer strength of node $j$ in layer $l$ is given by $\kappa_{jl} = k_{jl} + w_{jl}$. Finally, in the overall normalization, $\mu$ is the total strength of the adjacency tensor $\mathcal{A}$, given by $\mu = \frac{1}{2} \sum_{jl} \kappa_{jl}$. In Fig.~\ref{f:multilayer_schematic}, we show a schematic of a multilayer granular force network with evolving community structure. Importantly, the communities can persist across all layers and we can track their reconfiguration in terms of particle content and strength throughout the series of strain steps.

As in the formulation of $Q_{sing}$, the choice of null model in $Q_{multi}$ is an important one, particularly when considering systems with strict constraints on the allowed connectivity between nodes. For granular networks, we generalize the geographical null model to the multilayer regime. For community detection on the normal force network we use
\begin{equation}
\mathcal{P}^{n}_{ijl} = \langle f ^ {n} \rangle_{l} B_{ijl},
\label{eq:geog_null_norm}
\end{equation}
where  $\langle f ^ {n}  \rangle_{l}$ is the average of the normal component of the inter-particle forces at compressive step $l$ and $B_{ijl}$ is the contact matrix at compressive step $l$. For the tangential network, we take the null model to be
\begin{equation}
\mathcal{P}^{t}_{ijl} = \langle |f ^ {t}| \rangle_{l} \mathcal{B}_{ijl},
\label{eq:geog_null_tang}
\end{equation}
where  $\langle | f ^ {t} | \rangle_{l}$ is the average of the absolute value of the tangential component of the inter-particle forces at compressive step $l$, and $\mathcal{B}_{ijl}$ is the contact matrix at that step.

In each layer, we normalize the force network ($\mathcal{A}^{n}_{ijl}$ or ($|\mathcal{A}^{t}_{ijl}|$) by the mean inter-particle force in the corresponding layer. Thus, after normalization we have $\langle f ^ {n}  \rangle_{l} = 1$ in Eq.~\ref{eq:geog_null_norm} and $\langle | f ^ {t} | \rangle_{l} = 1$ in Eq.~\ref{eq:geog_null_tang}, for all $l$. This normalization ensures that the community structure is not purely driven by the final layer, which will have the largest total edge weight due to it being the most compressed.

\subsubsection{Choosing omega: flexibility as a measure of reconfiguration}
\label{s:choosing_omega}

The two parameters in the multilayer modularity quality function, $\omega$ and $\gamma$, must be chosen by the investigator. Recall that the structural resolution parameter, $\gamma$, regulates the size and number of detected communities; in this study, we use $\gamma_{l} \equiv \gamma = 1$ for all $l$. As can be seen from Eq.~\ref{eq:dynamic_modularity} and Eqs. \ref{eq:geog_null_norm} and \ref{eq:geog_null_tang}, the physical meaning of this value is that it selects for communities within each layer that have stronger than average force. The choice of $\omega$ is an active area of investigation; currently, there is no consensus in the literature on a single broadly-applicable method to determine the inter-layer coupling. In the present work, we make a physically informed choice. As described in the previous section, $\omega$ is in general a tensor that can take on different values between each layer or for different nodes. Since to our knowledge this is the first study on multilayer granular networks, we begin by taking the simplest case of a scalar inter-layer coupling, choosing $\omega$ to be the same for all particles and all pairs of layers, such that $\omega_{jlm} \equiv \omega$ for all $j, l, m$. However, it is important to point out that the inter-layer coupling could be different for different particles. For example, one could alternatively tune the relative value of $\omega$ for a given particle based on a particle property, such as total strength. Investigation of more complicated methods for choosing the inter-layer couplings may be an interesting direction for future work.

To proceed, it is necessary to understand the effect of $\omega$ on the community structure. There are two limiting cases which are relatively simple to grasp: when $\omega = 0$, there are no connections between layers of the adjacency tensor, and we therefore expect to recover the results of static community detection (Fig.~\ref{f:static_norm_comms}). At the other extreme, $\omega$ can be made large enough such that the strength of \textit{inter-layer} connections entirely overwhelms the strength of \textit{intra-layer} connections, resulting in completely consistent community structure across all compressive steps (that is, no observable changes; see Fig.~\ref{f:multilayer_comms_constant} in the Appendix for an example partition at large $\omega$). To understand what occurs in between these limiting cases, we consider a simple measure of network rearrangement called \textit{flexibility}, or $\Xi$, previously defined in \cite{Bassett:2011aa}. The flexibility of a single particle $i$, $\xi_{i}$, is defined as the number of times a particle changes in community allegiance across network layers, normalized by the number of possible changes. Mathematically,
\begin{equation}
\xi_{i} = \frac{g_{i}}{L-1},
\label{eq: node_flex}
\end{equation}
where $g_{i}$ is the number of times that the particle changes its community and $L$ is the total number of strain steps. The flexibility of the entire multilayer network is then given by the mean flexibility of all particles
\begin{equation}
\Xi= \frac{1}{N}\sum_{i}\xi_{i},
\label{eq: flex}
\end{equation}
where $N$ is the number of particles.

In order to choose a physically relevant value of the interslice coupling, we run 20 optimizations of multilayer community detection on the normal force network $\mathcal{A}^{n}$ for each packing, for several values of $\omega$ between 0 and 1, in steps of $\Delta \omega = 0.01$. Note that $\omega = 0.01$ corresponds to a coupling which is equal to $1/100$ of the mean edge weight in a given layer, due to the normalization procedure described in Sec.~\ref{s:multi_mod_max}. Due to the stochastic nature of modularity maximization \cite{Good:2010aa}, we avoid an inter-layer coupling close to zero, as the community structure is more likely to be influenced by noise in the algorithm. At each $\omega$, we compute $\Xi$ for all optimizations of a given packing, and then average over optimizations to obtain a single value of flexibility for the packing, $\langle \Xi \rangle_{opt}$. In what follows, we will denote averages over optimizations as $\langle \cdot \rangle_{opt}$, and averages first over optimization and then over packings with an overbar.

\begin{figure}[floatfix]
	\centering
	\includegraphics[width=0.6\columnwidth]{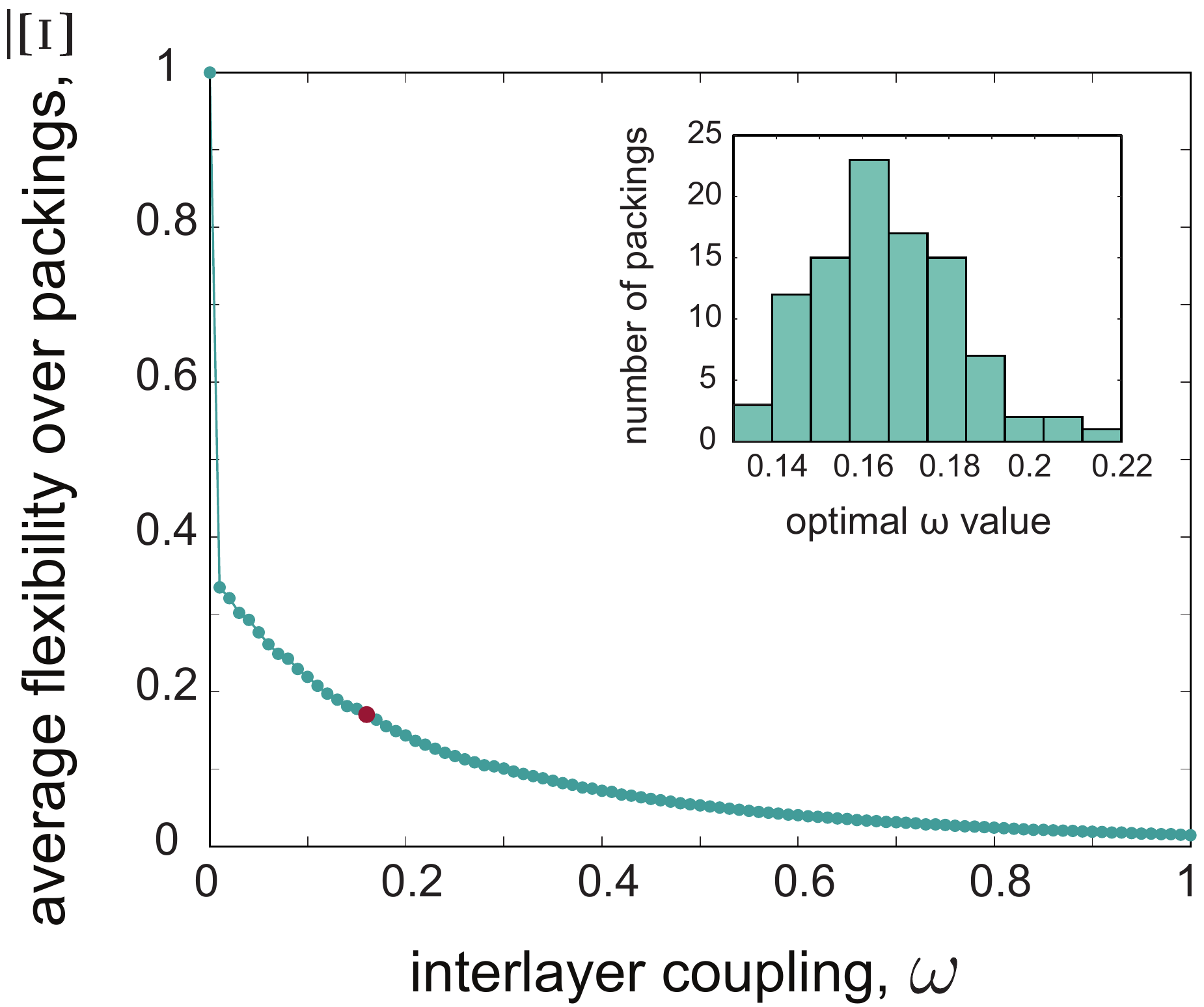}
	\caption{\label{f:flex_vs_omega} \textbf{Choosing an optimal inter-layer coupling.} The average network flexibility is governed by the value of the inter-layer coupling $\omega$. Each dot is the average of network flexibility over all experimental packings, $\overline{ \Xi }$, \emph{vs.} the inter-layer coupling, $\omega$. When $\omega = 0$, $\overline{\Xi}= 1$ and the results of static community detection are recovered; there is no persistent structure across compression. When $\omega = 1$, $\overline{ \Xi}  \approx 0$ and the particle contact forces within each layer are overwhelmed by the weight of the inter-layer coupling such that the community structure exhibits complete consistency throughout the compressive steps. At $\omega = 0.01$, $\overline{ \Xi_{max}} \approx 0.3349$. The optimal inter-layer coupling $\omega_{*} = 0.16$ is denoted by the red dot, and is chosen such that $\overline{ \Xi (\omega_{*})} \approx \overline{\Xi _{max}}$. This ensures a balance between intra- and inter- layer weights. The inset shows what the distribution of $\omega_{*}$ would be if the identical optimization were performed for each packing individually.}
\end{figure}

As shown in Fig.~\ref{f:flex_vs_omega}, we observe that the average flexibility $\overline{ \Xi }$ decreases smoothly as the inter-layer coupling $\omega$ increases. As expected, at $\omega = 0$, the network flexibility is unity, indicative of the fact that there is no consistency in community structure across layers. That is, at each step all particles are assigned to new communities which are independent from the community they were assigned to in the previous step. At $\omega = 1$, which corresponds to an inter-layer coupling of the same magnitude as the mean edge weight in each layer, the flexibility is approximately zero, and there are no changes in the community structure; instead, we observe complete consistency of community structure throughout layers.

Interestingly, at the first $\omega$ value away from zero, $\overline{ \Xi }$ sharply drops to $\approx$ 0.33. This steep decline can be explained by a recent result, which demonstrates that the case $\omega = 0$ is singular in the sense that even when $\omega = \epsilon$ with $0 < \epsilon \ll 1$, there will be at least some persistent community structure \cite{Bazzi:2016}. We thus take the value of $\overline{ \Xi}$ obtained from the first small step away from 0 at $\omega = 0.01$ to be the upper bound of network flexibility, $\overline{\Xi_{max}}$; from that point forward, $\overline{ \Xi }$ smoothly decreases to zero (see Appendix~\ref{a:choosing_omega} for further considerations into this point). To further characterize the behavior of the flexibilty \emph{versus} $\omega$ curve, we assessed whether the trend could be described as exponential decay, such that for each $\langle \Xi \rangle_{opt}$, we have $\langle \Xi \rangle_{opt} \approx Ae^{-\omega/\omega_{o}}$.  We find that the data can indeed be well approximated by an exponential. The $\omega_{o}$ for all packings fall in the range $ 0.26 \leq \omega_{o} \leq 0.40 $, and represent characteristic inter-layer couplings for the system.

At $\omega$ values too far into the exponential tail, the communities will not be sensitive to the structure present within a given layer, and at small values of $\omega$, temporal dependence becomes less important. We thus pick the optimal value of interslice coupling, $\omega_{*}$, to be the value such that $\overline{ \Xi(\omega_{*})}$ is closest to half of the maximum flexibility, $\overline{\Xi_{max}}$. This procedure yields a value of $\omega$ that balances the trade-off between the importance of intra-layer edges (particle contact forces) and persistent structure across network layers. In Sec.~\ref{s:null_models} we further validate this choice by comparing the community structure obtained at $\omega_{*}$ to three relevant null models, showing that the real network is distinguishable from the null models in each case.

Using the method described above, we find $\omega_{*} = 0.16$, (denoted by the red dot in Fig.~\ref{f:flex_vs_omega}), and we use this value in community detection for all packings and for both the normal and tangential force networks. The inset of Fig.~\ref{f:flex_vs_omega} shows what the distribution of $\omega_{*}$ would be if we were to optimize $\omega$ for each packing individually. We acknowledge that there are several other methods that could be used to determine an appropriate coupling, but here we have focused on a straightfoward method to choose a physically meaningful $\omega$, which yields intermediate values of network flexibility.

\section{Results \label{s:results}}

\subsection{Extraction of evolving mesoscale structure from the force network \label{s:community_extraction}}

\begin{figure*}[floatfix]
\includegraphics[width=\textwidth]{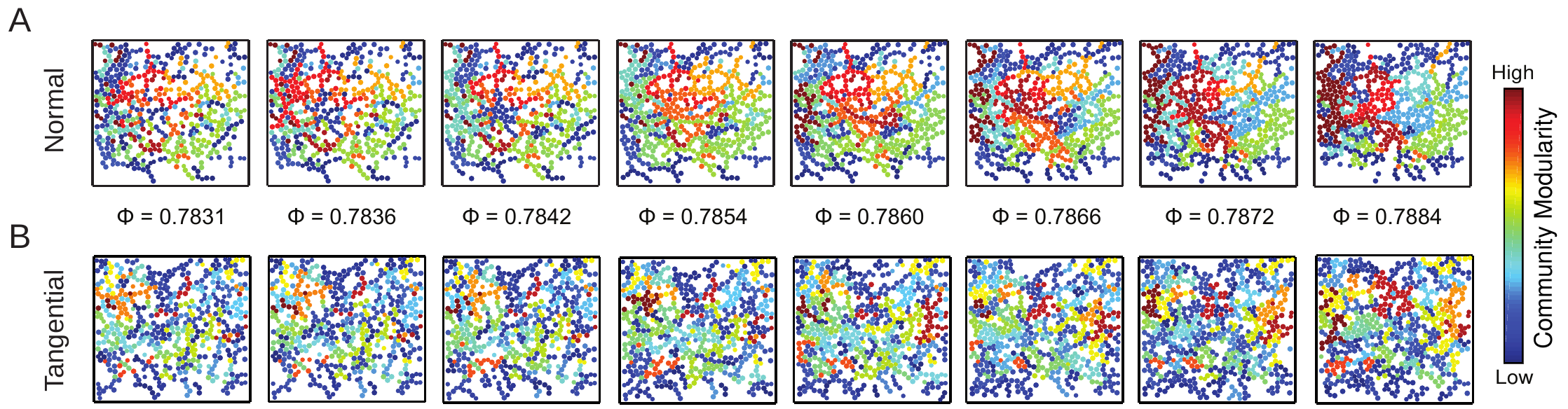}
\centering
\caption{\label{f:multilayer_communities} \textbf{Evolving community structure of a compressed granular packing.} For both the normal \emph{(A)} and tangential \emph{(B)} networks, particles are shown at their actual locations in physical space, and are colored according to their community assignment. Redder colors correspond to communities with higher multilayer modularity. The packing fraction increases from left to right, and only the steps with $\Phi > \Phi_{J}$ are shown. Visually, the normal force communities appear to become more compact throughout compression, and the emergence of force chain like structure is observed. The tangential network forms noticeably smaller groups of particles. We quantify these physical differences, as well as differences in network reconfiguration, throughout the main text.}
\end{figure*}

To extract pressure dependent particle assemblies (communities) from the multilayer force networks of each experimental run, we maximize multilayer modularity (Eq.~\ref{eq:dynamic_modularity}). We consider both the normal and tangential force networks separately, with $\mathcal{A}_{ijl}$ defined as in Sec.~\ref{s:multi_network}, and $\mathcal{P}_{ijl}$ given by Eq.~\ref{eq:geog_null_norm} (or \ref{eq:geog_null_tang}). As determined in the previous section, in both cases we use $\gamma = 1$ and $\omega = \omega_{*} = 0.16$. For each particle configuration, we carry out 20 maximizations of the multilayer modularity to obtain an ensemble of partitions, each with their respective value of $Q_{\mathrm{multi}}$. 
	
By the nature of modularity maximization and our choice of null model, the resulting communities correspond to spatially localized, mesoscale groups of particles that display collective organization throughout the compression process. In Fig.~\ref{f:multilayer_communities} we show an example of the community architecture at each step above $\Phi_{J}$ detected from the normal force network (\emph{A}) and tangential force network (\emph{B}), of a particular experimental configuration. In both cases, communities are colored according to their multilayer modularity value, $Q_{multi}$. Importantly, the same color at each strain step corresponds to the same community, to provide a visual sense of linking particle assemblies continuously throughout compression. Unlike in the single layer modularity maximization, the communities here can persist across layers, providing a means to directly examine network evolution and reconfiguration. 

It is important to point out that the structure uncovered with multilayer community detection is representative of the system as a whole across all layers, and not necessarily of the static force chain structure of individual layers. Because of this, it is not required that a community in a given layer consists of only physically connected particles; rather communities in a given layer are dependent on the structure of the entire temporal network. In this way, the communities embody changing network architecture, and we can thus observe the break up and coalescence of communities across the applied compressive steps.  By eye, the mesoscale organization appears to become more compact, and we observe the emergence of force chain-like organization as the packing fraction increases. As expected for a compression process, these structural patterns are most clearly evident for the normal force network. The communities from the tangential network appear visually smaller than those extracted from the normal forces. In the following sections, we study the reconfiguration and physical properties of this evolving architecture, and quantify differences between the normal and tangential mesostructures.

\subsection{Characterization of mesoscale reconfiguration}
\label{s:reconfig}

We characterize the evolution of the multilayer community structure using two diagnostics: network flexibility, $\Xi$ (Eq.~\ref{eq: flex}), and community stationarity, $\zeta_{c}$. Recall that the network flexibility quantifies the amount of reconfiguration in the system at the particle level, determined by the fraction of steps over which a particle changes its community allegiance. As a second measure of structural reconfiguration we also consider the community stationarity \cite{Gergely:2007aa,Bassett:2013aa}, which measures the consistency of particle content in each community throughout compression.

\begin{figure}[floatfix]
\includegraphics[width=\columnwidth]{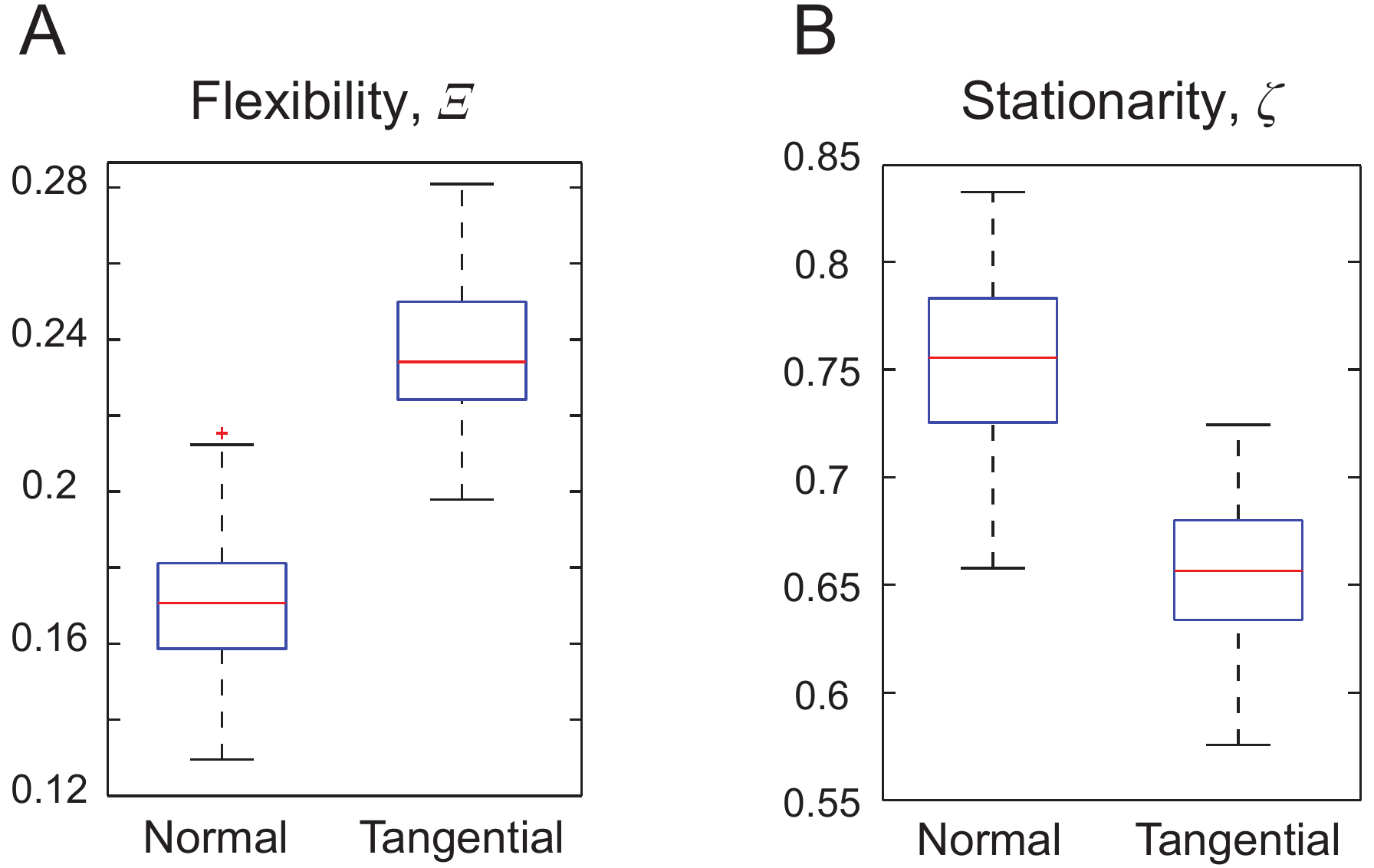}
\centering
\caption{\label{f:dynamic_stats} \textbf{Measures of evolving community structure for the normal and tangential force networks.} The structure and reconfiguration of the community structure of the normal and tangential force networks is characterized by network flexibility $\Xi$ \emph{(A)} and network stationarity $\zeta$ \emph{(B)}. The boxplots are constructed by averaging statistic values over optimization for each experimental packing. The red line then denotes the median over packings and the edges correspond to the $25^{th}$ and $75^{th}$ percentiles. For the same inter-layer coupling, the normal and tangential networks have statistically different behavior in terms of these measures of network rearrangement.}
\end{figure}

To define stationarity, we begin by writing the autocorrelation $J(c_{l}, c_{l+m})$ between a given community at layer $l$, $c_{l}$, and the same community at layer $l+m$, $c_{l+m}$, as
\begin{equation}
J(c_{l}, c_{l+m})= \frac{|c_{l} \cap c_{l+m}|}{|c_{l} \cup c_{l+m}|}
\label{eq: jaccard}
\end{equation}
where $|c_{l} \cap c_{l+m}|$ is the number of particles present in community $c$ at strain step $l$ that are \textit{also} present in community $c$ at step $l+m$, and $|c_{l} \cup c_{l+m}|$ is the number of distinct particles present in community $c$ at strain step $l$ \textit{or} step $l+m$. Then if $l_{i}$ is the layer in which community $c$ first appears, and $l_{f}$ is the layer in which it last appears, the stationarity of community $c$ is
\begin{equation}
\zeta_{c} = \frac{\sum_{l = l_{i}}^{l=l_{f}-1} J(c_{l}, c_{l+1})} {l_{f} - l_{i}}.
\label{eq: station}
\end{equation}
In this way, communities that experience large changes in their particle content over consecutive compressive steps will have larger values of $\zeta_{c}$ than communities with more consistent structure. The average stationarity of a multilayer network is obtained by taking the mean of $\zeta_{c}$ over all $n_c$ communities:
\begin{equation}
\zeta = \frac{1}{n_{c}} \sum_{c} \zeta_{c},
\label{eq: avg_station}
\end{equation}
where we exclude \textit{singleton} communities, which only contain one particle in all layers. In Fig.~\ref{f:dynamic_stats} we show boxplots over the particle configurations of $\Xi$ and $\zeta$ for the normal and tangential force networks. We have averaged these diagnostics over the 20 optimizations, to obtain a mean value of $\langle \Xi \rangle _{opt}$ and $\langle \zeta \rangle_{opt}$ for each experimental packing. 

We test whether the reconfiguration of the normal and tangential force networks can be distinguished by performing non-parametric permutation tests on the flexibility and stationarity values. For the network of normal forces, there are 97 values of the flexibility $\langle \Xi ^{n}\rangle _{opt}$ and stationarity $\langle \zeta^{n}\rangle_{opt}$ (one for each laboratory configuration). Repeating the same protocol for the tangential forces yields another set of values $\langle \Xi ^{t} \rangle _{opt}$ and $\langle \zeta^{t}\rangle_{opt}$. We calculate the difference in the means of these two distributions, and test whether that difference is greater than expected in the null distribution created by re-assigning statistics uniformly at random to the two groups: ``normal'' and ``tangential''. Using this test, we find significant differences in the means (over optimizations and then packings) of both statistics, with both $p$-values less than $1 \times 10^{-3}$. In particular, $\overline{\Xi^{n}} < \overline{\Xi^{t}}$ and  $\overline{\zeta^{n}}  > \overline{\zeta^{t}}$. From this finding, we conclude that at the same value of inter-layer coupling, the multilayer network of normal forces tends to exhibit less reconfiguration during compression than the network of tangential forces. This sensitivity to differences in the evolution of two related but distinct force networks suggests that our method may be more broadly applicable. For example, it could be used to test for differences and classify different types of granular systems composed of varying particle materials, shapes, or sizes. 

\subsection{Reconfiguration of network architecture during compression: a null model comparison} 
\label{s:null_models}

\begin{figure*}[floatfix]
\includegraphics[width=\textwidth]{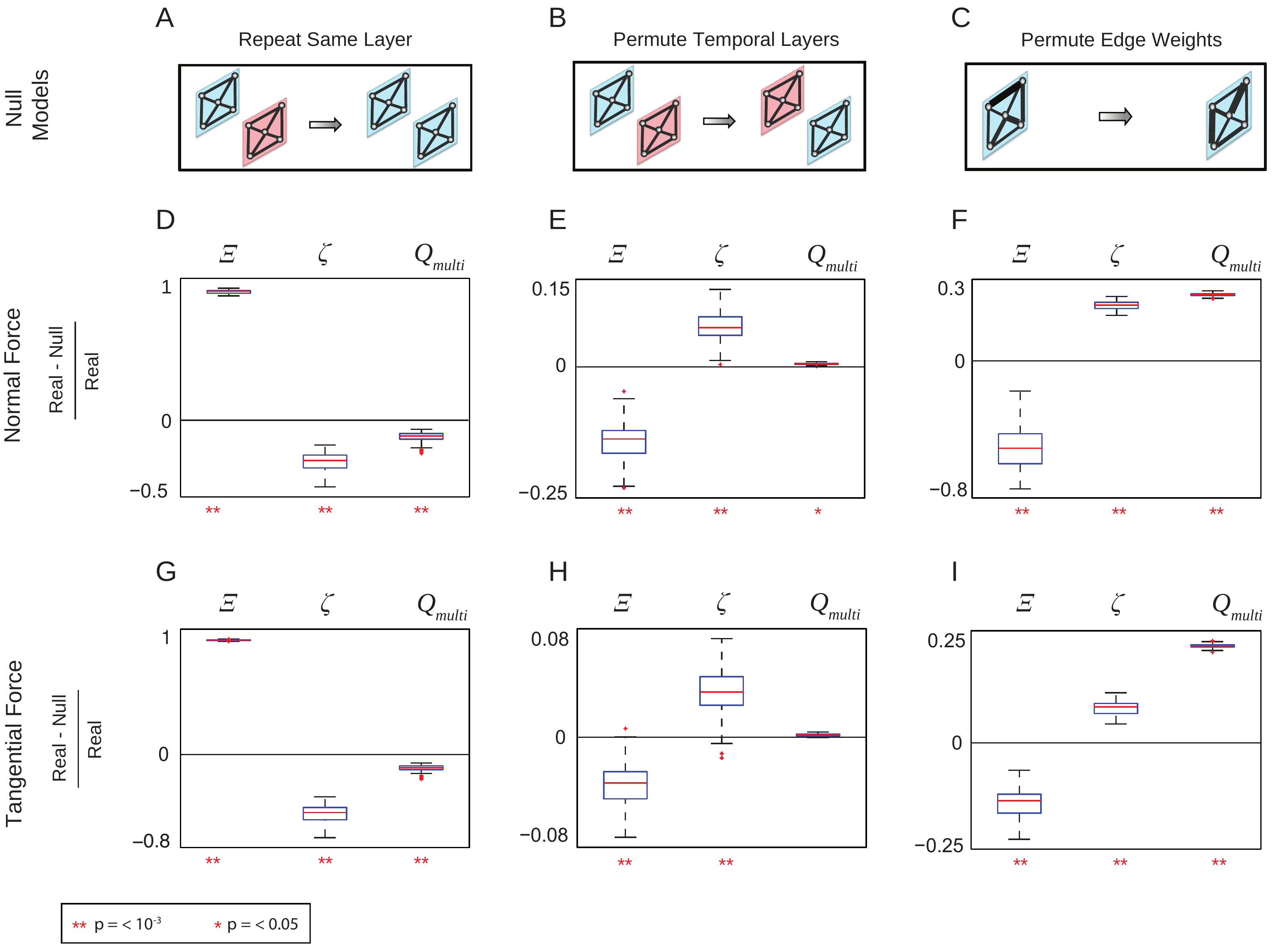}
\centering
\caption{\label{f:null_models} \textbf{A comparison of granular network evolution to three null networks.} The community structure of the multilayer granular system is compared to three relevant null models (top row), using the quantities $\Xi$, $\zeta$, and $Q_{multi}$. For each null model, the boxplots show the normalized difference between the real and null network diagnostics for all experimental configurations, and for the normal (second row) and tangential (third row) components. Significant differences between the real and null networks exist when the boxplots are above or below the zero line. The $p$-values obtained from permutation testing are shown beneath each box to quantify the significance of results. (If no $p$-value is shown, the difference is not significant). \emph{(A)} A null model formed by repeating the same layer for all steps, $\mathcal{A}^{repeat}$. The values $\Xi$, $\zeta$, and $Q_{multi}$ are statistically different between the real and null networks for the normal \emph{(D)} and tangential \emph{(G)} components. \emph{(B)} A null model, $\mathcal{A}^{temporal}$, constructed by permuting the temporal layers uniformly at random. The values $\Xi$, $\zeta$, and $Q_{multi}$ are statistically different between the real system and null model for the normal forces \emph{(E)}, and $\Xi$ and $\zeta$ are statistically different between the real system and null model for the tangential forces \emph{(H)}. \emph{(C)} The third null model, $\mathcal{A}^{edges}$, is built by permuting the edge weights uniformly at random, while maintaining the original contact topology and temporal ordering. The values $\Xi$, $\zeta$, and $Q_{multi}$ are statistically different between the real and null networks for both the normal \emph{(F)} and tangential \emph{(I)} components.}
\end{figure*}

Perhaps more important than absolute values of network measures is the question of whether or not the network evolution observed in the real, physical system is significantly different than what is expected from relevant null models, and whether or not our model is sensitive to these differences. In this section, we demonstrate that the multilayer community structure of the compressed granular configuration is indeed distinct from three null models with respect to the diagnostics defined previously ($\Xi$, and $\zeta$). Here, we additionally consider the multilayer modularity, $Q_{multi}$. (Recall that $Q_{multi}$ is a general measure of how well the network can be partitioned into densely interconnected groups of particles throughout the compression process, with respect to the physically appropriate geographical null model). To ensure a fair comparison, we test the real force networks from each of the 97 experimental realizations against null models that are built using the force information from the same experimental run. Furthermore, we analyze the normal and tangential forces separately. To determine statistical significance, we perform permutation; assignments of statistic values to the two groups, ``real network" or ``null network", are permuted uniformly at random to construct a null distribution of expected differences between the two distributions.

In the first case, we impose the elimination of steady perturbation on the system, comparing the real multilayer force networks to null models built by setting all layers to be equal (Fig.~\ref{f:null_models}\emph{A}). In particular, we construct a null network, $\mathcal{A}^{repeat}$, by repeating the real force network at a constant layer $s$, $L$ times (recall that $L$ is the total number of layers). Carrying out this process for all such constant layers yields a set of $L$ null networks for each packing. We run community detection 20 times for each null network, and compute the network diagnostics ($Q_{multi}^{null}$, $\Xi^{null}$, and $\zeta^{null}$) for each trial. We average quantities first over the optimization realizations, and then over the different networks. Physically, this simple null model serves as a control that allows us to assess the implications and consequences of compression on the detected community structure. Since there are no changes in topological structure or edge weights from one layer to the next in the synthetic networks, we expect all changes in community structure to be due to noise. This baseline network reconfiguration should be much less than the reconfiguration that occurs in the real networks, which encode the compression procedure. Indeed, for both the normal and tangential force networks, we see that the null models have significantly lower flexibility, $\Xi^{null} < \Xi^{real}$, and higher stationarity, $\zeta^{null} > \zeta^{real}$, than the compressed system (see Fig.~\ref{f:null_models}\emph{A,D,G}). Note also that the modularity of the null model is greater than that of the true networks, which is also expected since the null models will be partitioned into highly temporally consistent community structure.

We next compare the real system to a null model, $\mathcal{A}^{temporal}$, constructed by permuting the temporal layers (strain steps) of the real networks uniformly at random (Fig.~\ref{f:null_models}\emph{B}). In the experimental protocol, compression is applied systematically, in small and always increasing steps. The temporal null model considered here effectively eradicates this regularity in the layer ordering. The expectation is thus that the real networks will have less reconfiguration than the scrambled model. For each experimental packing, we create 20 null networks built from different random permutations of the layers, and run 20 optimizations of the Louvain algorithm for all permutations. As before, we compute $\Xi^{null}$, $\zeta^{null}$, and ${Q}^{null}_{multi}$, and average the results first over optimizations and then over permutations. In Fig.~\ref{f:null_models}\emph{E,H} we compare the real system to the null model, finding that our method is again sensitive to differences in the evolution of the real and synthetic networks. For both the normal and tangential force networks, $\Xi^{null} > \Xi^{real}$, and $\zeta^{null} < \zeta^{real}$ implying more steady and regular progression of community structure in the real system. In addition, we observe a slight decrease in modularity in the temporally permuted normal force network, suggesting that the real compression protocol yields stronger multilayer community structure. The modularity of the tangential forces is less affected by temporal scrambling than that of the normal forces. We also find that the real system can be distinguished from the temporal null model with respect to an alternative measure of reconfiguration known as \textit{promiscuity}, $\Psi$ \cite{promiscuity_code}. See Appendix~\ref{a:promiscuity} for a definition of this statistic and the results of the null model analysis.

In the final null model (Fig.~\ref{f:null_models}\emph{C}), we consider the spatial distribution of forces throughout the system. We construct a null model, $\mathcal{A}^{edges}$, by permuting the edge weights uniformly at random within each layer while maintaining the original contact topology and ordering of slices (for a related but distinct null model, see \cite{Smart:2008aa}). It is known that the organization of inter-particle forces is crucial for the stability of granular packings. This fact is manifested in force chains, branching groups of particles that bear the majority of the load in the system. Therefore, for the multilayer model to be useful, it should not be agnostic to the pattern of forces present in physically realizable systems. For each of the 97 configurations, we form an ensemble of null networks by permuting the forces uniformly at random within each layer 20 different times, and then optimize modularity 20 times for each permuted network. As before, this is done separately for the normal and tangential forces. In this case, we expect the synthetic networks to display more reconfiguration and less modular structure than the physical networks. We observe that the diagnostics of multilayer community structure are highly distinguishable between the real multilayer force networks and the null model (Fig.~\ref{f:null_models}\emph{F,I}). In particular, the force-permuted networks exhibit more flexibility, $\Xi^{null} > \Xi^{real}$, less stationarity $\zeta^{null} < \zeta^{real}$, and decreased modularity $Q_{multi}^{null} < Q_{multi}^{real}$ for both the normal and tangential components than expected in the null model. These results confirm our hypothesis that the evolution of network structure is less variable and undergoes less reorganization in the real system, and that the modularity stronger. In the Appendix~\ref{a:promiscuity}, we additionally show that these results hold for the promiscuity.

The findings presented in this section crucially demonstrate that the multilayer network model and community detection are sensitive to differences in the structural evolution of the compressed system compared to three relevant null models, with respect to $\Xi$, $\zeta$, and $Q^{multi}$. In particular, we can distinguish between the evolution of the granular networks and null models with consistent topology and weights, scrambled temporal layers, or scrambled edge weights (but same topological structure). Even more importantly, the differences in the reconfiguration agree with what is expected from a physical standpoint, making evident the powerful utility of this framework.

\subsection{Physical properties of evolving mesostructures}

In the previous section, we examined three measures to quantify the community structure of the multilayer granular network. However, the diagnostics we considered did not directly describe the evolution of physical properties of the mesoscale organization. We turn now to a physical description of the network architecture, and define measures to quantify both the strength and geometry of community structure throughout compression. We then ask whether the physical properties of mesoscale particle assemblies can be related to the measures of structural reconfiguration previously defined.

\subsubsection{Evolution of community characteristics throughout compression}
\label{s:physical_properties}

We characterize the physical nature of community structure with three measures: \textit{size}, \textit{strength}, and \textit{sparsity}, and examine how these quantities evolve throughout compression. Similarly to the previous section, we consider and compare both the normal and tangential force network. The size $s^{c}_{l}$ of community $c$ at strain step $l$ is simply the number of particles within the community at that step. We denote the strength of a community at layer $l$ as $\sigma^{c}_{l}$, and define it to be the average amount of force (normalized by the mean force in the layer) on a particle in community $c$ from intra-community contacts. Mathematically,
\begin{equation}
\sigma^{c}_{l} = \frac {1} {s^{c}_{l}} \sum_{i,j  \in c_{l}} \mathcal{A}_{ijl},
\label{eq: comm_strength}
\end{equation}
where $\mathcal{A}$ is either the normalized normal or tangential force network. Finally, we consider a measure of spatial compactness, which we term the community sparsity, $\eta^{c}$. The sparsity of a community is closely related to the \textit{hull ratio}, which has been defined and used to quantify the geometric arrangement of compressed particle assemblies \cite{Huang:2015}. The hull ratio $h^{c}_{l}$ of a community at strain step $l$ can be understood as the ratio of the area $a$ of particles in the community to the area of the convex hull of the group $a_{hull}$, such that
\begin{equation}
h^{c}_{l} =  \frac {\sum_{i \in c_{l}} a_{i}} {a_{hull}}
\label{eq: comm_hull}
\end{equation}
where the area of particle $i$ is $a_{i} = \pi r_{i}^{2}$, with $r_{i}$ equal to the particle radius. We then take the community sparsity at layer $l$ to be
\begin{equation}
\eta^{c}_{l} = 1 - h^{c}_{l}.
\label{eq: comm_sparsity}
\end{equation}
With this definition, small values of $\eta$ correspond to spatially dense groups of particles, and high values of $\eta$ correspond to sparse particle configurations.

We now use each of these physical characteristics to quantify the evolution of the mesoscale community structure throughout the compression process. Given a partition of particles into communities, we compute $s^{c}_{l}$, $\sigma^{c}_{l}$, and $\eta^{c}_{l}$ at each strain step $l$, for all communities except those which have $s^{c}_{l} = 1$ for all $l$ (i.e., they are always singletons). We then define $s_{l}$, $\sigma_{l}$, and $\eta_{l}$ to be the average over all communities present in layer $l$, 
\begin{subequations}
	\begin{equation}
		s_{l} = \frac{1}{n_{c_{l}}} \sum_{c_{l}} {s^{c}_{l}} 
	\end{equation}
	\begin{equation}
		\sigma_{l} = \frac{1}{n_{c_{l}}} \sum_{c_{l}} {\sigma^{c}_{l}}
	\end{equation}
	\begin{equation}
		\eta_{l} = \frac{1}{n_{c_{l}}} \sum_{c_{l}} {\eta^{c}_{l}} 
	\end{equation}
\end{subequations}
where $n_{c_{l}}$ is the number of communities present in layer $l$. Repeating this process for all community detection optimizations and all particle configurations, we form representative curves of each physical quantity as a function of strain step by averaging the measures defined above first over optimizations and then over experimental configurations. We denote the final averaged physical quantities for size, strength, and sparsity as $\langle s \rangle$, $\langle \sigma \rangle$, and $\langle \eta \rangle$, respectively.

\begin{figure*}
	\centering
	\includegraphics[width = \textwidth]{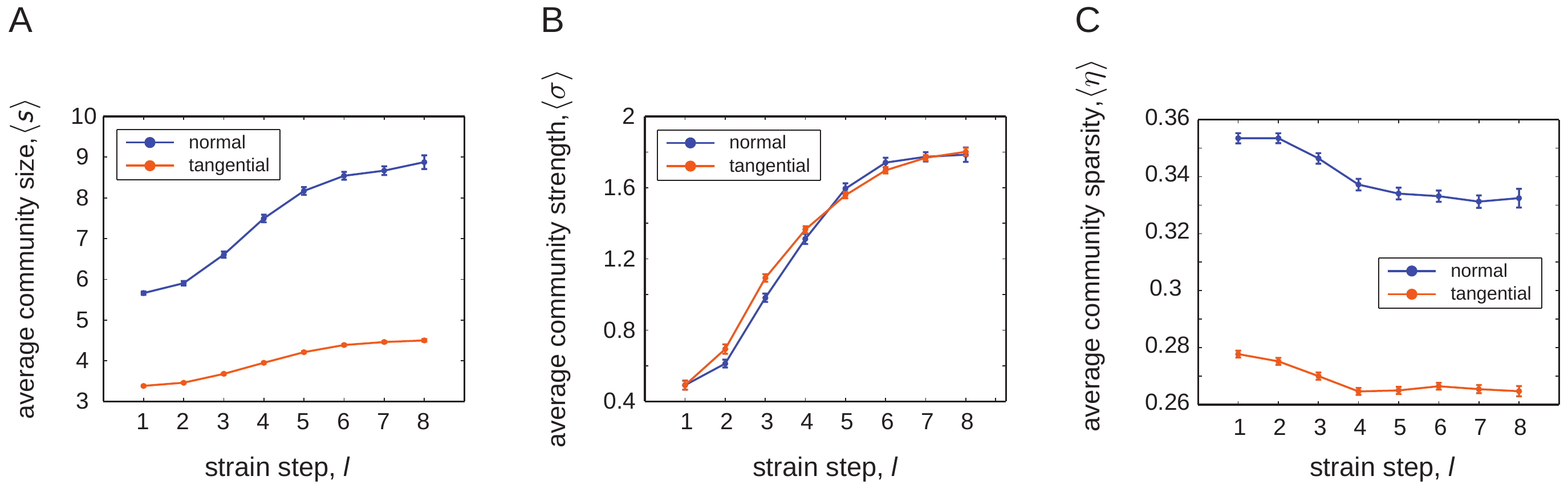}
	\caption{\label{f:comm_physical} \textbf{The evolution of physical characteristics of community structure throughout compression}. Community structure is characterized by size, strength, and sparsity for the normal (blue) and tangential (orange) networks. \emph{(A)} The average community size $\langle s \rangle$ of the normal and tangential networks increases with strain step, but the tangential force communities tend to be smaller than the normal force communities. \emph{(B)} The community strength $\langle \sigma \rangle$ measures the intra-community forces, and also grows as a function of strain step for both the normal and tangential networks. \emph{(C)} The sparsity $\langle \eta \rangle$ quantifies the spatial density of community structure. The tangential force communities tend to be more dense than the normal force communities. In all plots, error bars correspond to the standard error of the mean over packings.}
\end{figure*}

We first observe that the size of the mesoscale structure increases smoothly as a function of the strain steps, (Fig.~\ref{f:comm_physical} \emph{A}), suggesting that the scale of the mesostructure organization grows with compression. However, the normal force communities are noticeably larger in size, and undergo relatively more growth throughout strain steps than the communities from the tangential network. These differences imply that the network of normal forces exhibits collective organization on a larger scale than that of the tangential force network; and that the normal force network responds differently to increasing pressure. Some of these features are partially recognizable by eye in comparing the community structure in Fig.~\ref{f:multilayer_communities}\emph{A} and Fig.~\ref{f:multilayer_communities}\emph{B}. We also find that the community strength increases smoothly over the applied strain steps, for both the normal and tangential components. This behavior signifies the mesoscale architecture becoming more strongly connected throughout compression, which agrees with the physical expectation. Note that the average community strength $\langle \sigma \rangle$ was computed on the normalized networks (see Sec. \ref{s:multi_mod_max}), which explains the similar scale between the normal and tangential curves. Finally, we observe a slight decrease in community sparsity across strain steps, especially during the beginning stages. In addition, the tangential network displays lower sparsity than that of the normal force network, implying more compact tangential community structure. This feature is likely tied to the smaller size of tangential communities, which constrains the set of possible spatial arrangements of the particles within a community.

\subsubsection{Trends in physical characteristics are diverse} 
\label{s:phys_trends}

In addition to quantifying the average behavior of mesoscale architecture, it is also important to investigate the behavior of individual communities throughout compression. Though it is possible that all communities progress similarly to the average behavior of the system (for example, coalescing to create communities of increasing size and strength, but decreasing sparsity), this does not have to be the case. We find, in fact, that the situation is quite the opposite; at the level of single communities, the evolution of physical structure varies greatly. We demonstrate this in a simple way. First, we identify the number of communities that exhibit linear trends with respect to size, strength, or sparsity as a function of strain step. Tables \ref{t:normal_trends} and \ref{t:tangential_trends} show the results of this analysis for the normal and tangential networks. We observe that the majority of mesostructures do \textit{not} exhibit consistent and predictable linear trends in terms of their physical properties throughout the compression process. While some linear tendencies are much more likely to occur than others (for example, increasing strength and decreasing sparsity), the behavior of many communities cannot be characterized by a simple linear relationship. This surprising results highlights the important diversity of mesoscale structural evolution.

\begin{table}[h]
\begin{centering}
\begin{tabular}{p{3cm} p{3cm} l}
\hline \hline
Trend & Strength, $\sigma$ & Sparsity, $\eta$ \\
\hline
Increasing & 32.8 $\pm$ 0.6 & 12.2 $\pm$  0.3 \\
Decreasing & 3.8 $\pm$ 0.3 & 42.3 $\pm$ 0.8 \\
\hline \hline
\end{tabular}
\caption{\label{t:normal_trends} Percentages of communities from the network of normal forces that exhibit linear trends with respect to strength $\sigma$ or sparsity $\eta$ throughout compression. The numbers reported correspond to averages over optimizations and packings, and errors are the standard errors of the mean.}
\end{centering}
\end{table}

\begin{table}[!h]
\begin{centering}
\begin{tabular}{p{3cm} p{3cm} l}
\hline \hline
Trend & Strength, $\sigma$ & Sparsity, $\eta$ \\
\hline
Increasing & 18.3  $\pm$ 0.3 & 14.9 $\pm$ 0.3 \\
Decreasing & 2.3  $\pm$ 0.1 & 28.1  $\pm$  0.6 \\
\hline \hline
\end{tabular}
\caption{\label{t:tangential_trends} Percentages of communities from the network of tangential forces that exhibit linear trends with respect to strength $\sigma$ or sparsity $\eta$ throughout compression. The numbers reported correspond to averages over optimizations and packings, and errors are the standard errors of the mean.}
\end{centering}
\end{table}

\begin{figure}[!h]
	\centering
	\includegraphics[width=0.8\columnwidth]{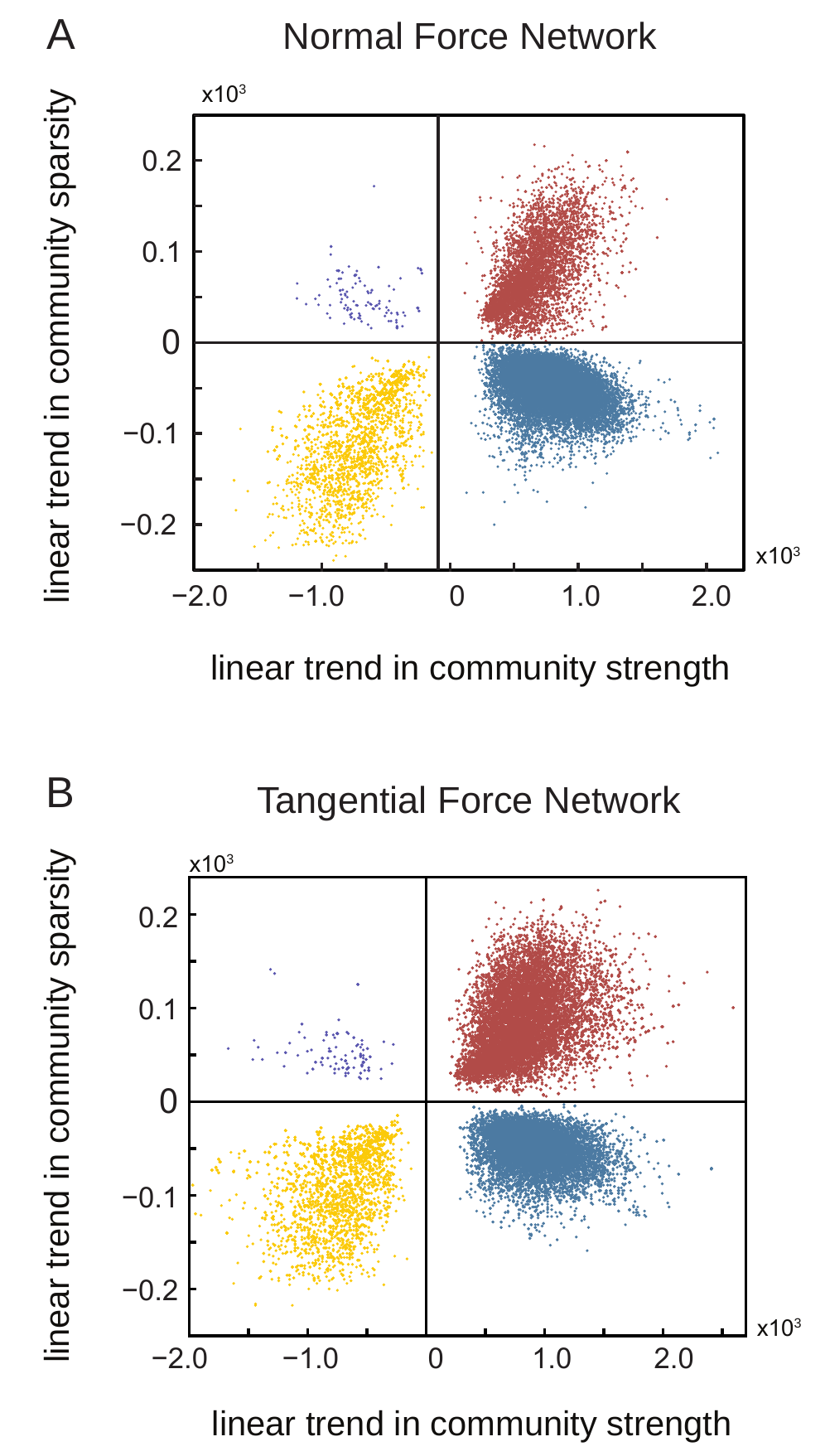}
	\caption{ \label{f:force_hull_trend} \textbf{Scatter plots demonstrating the diversity of network evolution}. The relationship between the slopes of communities that exhibit linear trends with respect to strength and sparsity. For both the normal \emph{(A)} and tangential \emph{(B)} force networks, all but the upper left quadrant are quite populated, pointing to the diversity in the evolution of community structure throughout the compression process.}
\end{figure}

Next, we ask if and how the set of communities which \textit{do} have linear behavior with respect to a given physical property, are related to each other. In Fig.~\ref{f:force_hull_trend}, we plot the slope of the linear regression fit of sparsity \emph{vs.} the slope of strength, for each community in all optimizations and experimental packings.  Again, the scatter plots point to the variation of mesostructure development, as all quadrants (with the exception of the upper left) are significantly filled in. These data support the notion that communities may coalesce, disband, or become increasingly branch-like, and each of these behaviors is observable as the packing rearranges under compression. 

To quantify the co-dependence of the two statistics, we first find the number of communities (i.e. the intersection) that fall within each quadrant for each optimization and packing. For example, if $\sigma_{\uparrow}$ are the communities with linearly increasing strength and $\eta_{\uparrow}$ are those with linearly increasing sparsity, then we compute the number of communities that satisfy $\sigma_{\uparrow} \cap \eta_{\uparrow}$ (upper right quadrant of the scatter plots). Then to determine how often increasing strength ($\sigma_{\uparrow}$) occurs with increasing sparsity ($\eta_{\uparrow}$), for example, we normalize the intersection by the total number of communities with $\sigma_{\uparrow}$. Conversely, if we want to know the percentage of communities with increasing sparsity ($\eta_{\uparrow}$) that also have increasing strength ($\sigma_{\uparrow}$), then we would instead divide the intersection by the number of communities in $\eta_{\uparrow}$. Table \ref{t:trend_together} in Appendix~\ref{a:table_trend} shows the percentages for each of the possible combinations. The strongest relationship occurs between communities with $\sigma_{\downarrow}$ and $\eta_{\downarrow}$. In this case, we find that if a community has linearly decreasing strength with strain step, then it is likely to also be more compact (but note that not many communities decrease in strength in the first place). These results may be due to the community losing particles and thus becoming more dense (on average for the normal (tangential) network, $98.6 \%$  ($96.6 \%$) of communities with linearly decreasing strength and sparsity also have linearly decreasing size). The nearly empty upper left quadrants are consistent with this relationship as well; communities with decreasing strength rarely become more spatially spread out throughout compression. We also find that on average, more than half of the communities with increasing sparsity also have increasing strength, which is likely due to the community gaining particles, thus allowing it to take on configurations which are more spatially extended (on average for the normal (tangential) network, $93.9 \%$  ($91.7 \%$) of communities with linearly increasing strength and sparsity also have linearly increasing size).

\subsection{Linking physical properties to network reconfiguration}

Thus far, we have independently characterized the evolution of multilayer community structure using notions of particle rearrangement (flexibility and stationarity), and using physical quantities, (size, strength, and sparsity). We now attempt to link these two ideas together, asking whether network reconfiguration can be related to physical aspects of the network.

\subsubsection{Local reconfiguration}
\label{s:local_reconfiguration}

We first investigate the relationship between particle flexibility, $\xi$ (Eq.~\ref{eq: node_flex}) and the inter-particle force, $f$. Recall that $\xi$ is a measure of \textit{local} reconfiguration in that it is defined for a single particle, but it is determined from the \textit{mesoscale} community structure of the network. For every multilayer community partition (20 for each particle configuration), we compute the flexibility of each particle as given in Eq.~\ref{eq: node_flex}, and average these values over partitions. This yields a single value of flexibility $\xi_{i}$ for the $i^{th}$ particle in a given experimental run. We do this for the normal and tangential force networks separately.

Our first finding is that flexibility $\xi$ is strongly correlated with the average force on a particle throughout strain steps $\langle f \rangle_{\phi}$, as well as the average absolute change in force on the particle $\langle |\Delta f| \rangle_{\phi}$ between consecutive strain steps. This result holds for both the normal and tangential components. For the $i^{th}$ particle, the average change in force $\langle |\Delta f^{i}| \rangle_{\phi}$ is given by

\begin{equation}
\langle |\Delta f^{i}| \rangle_{\phi} = \frac {1} {L - 1} \sum_{l = 1}^{l = L - 1} |f_{l+1}^{i} - f_{l}^{i}|,
\label{eq: change_in_force}
\end{equation}
where $f^{i}_{l}$ is the total force on the $i^{th}$ particle in layer $l$, determined from the adjacency tensor as $f^{i}_{l} = \sum_{j} \mathcal{A}_{ijl}$.

\begin{figure}[floatfix]
\includegraphics[width=\columnwidth]{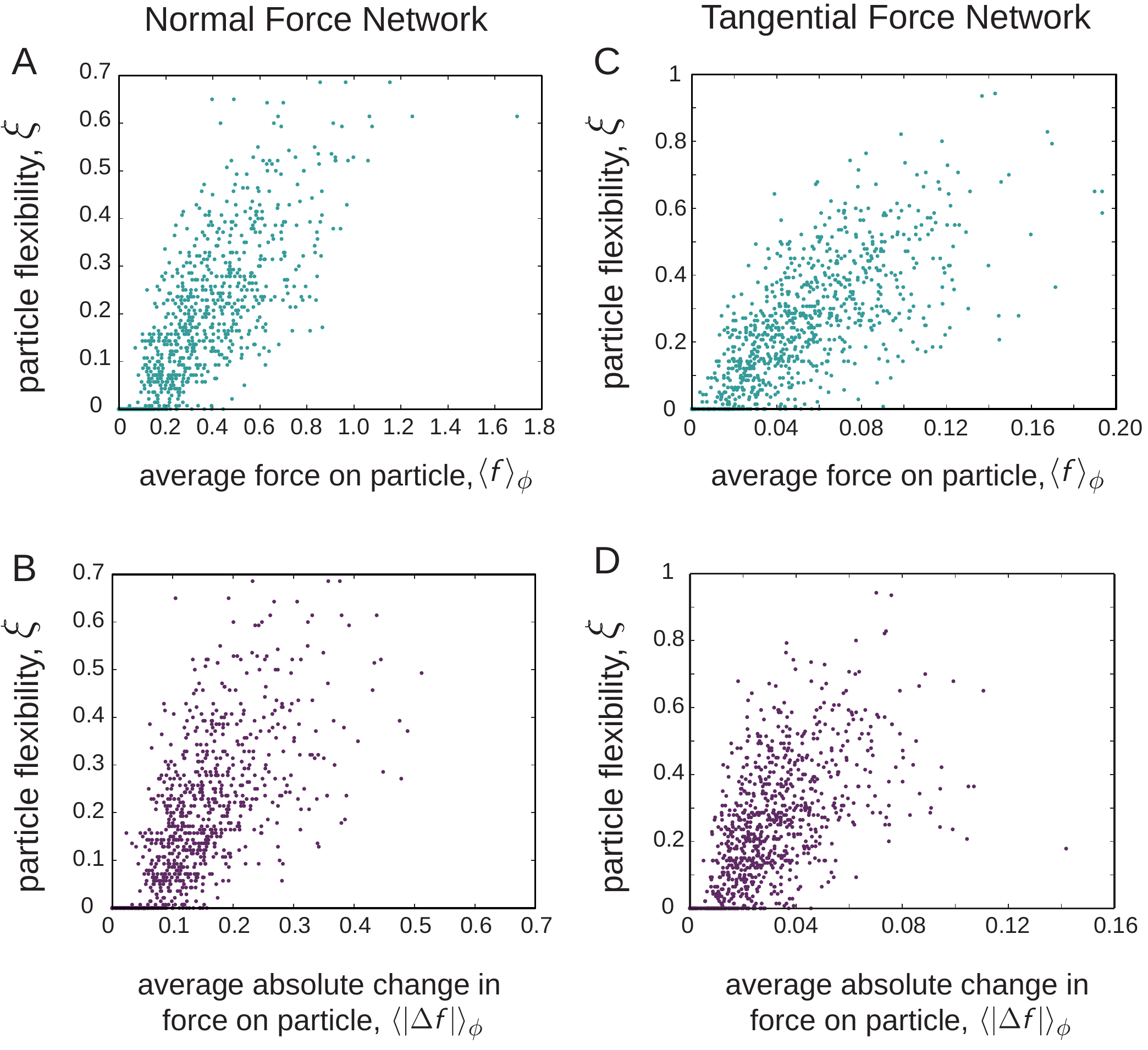}
\centering
\caption{\label{f:force_vs_flex} \textbf{Force drives local reconfiguration}. \emph{(A, C)} Scatter plots of particle flexibility $\xi$ \emph{vs.} the average force on a particle across compression $\langle f \rangle_{\phi}$ for a sample packing. In both the normal and tangential networks, there is a strong, positive Spearman correlation between the two quantities. \emph{(B, D)} Scatter plots of particle flexibility $\xi$ \emph{vs.} the average absolute change in force on a particle across compression $\langle | \Delta f | \rangle_{\phi}$ for a sample packing. In both the normal and tangential networks, there is a strong, positive Spearman correlation between the two quantities. See Fig.~\ref{f:Spearman_flex} for the distribution of correlations for each packing).}
\end{figure}

In particular, we observe that particles with high flexibility $\xi$ also tend to have high values of average force $\langle f \rangle_{\phi}$ and average change in force $\langle |\Delta f| \rangle_{\phi}$. In Fig.~\ref{f:force_vs_flex}, we plot $\xi$ \emph{vs.} $\langle f \rangle_{\phi}$ for each particle using the normal \emph{(A)} and tangential \emph{(C)} force networks of one experimental configuration.  Panels \emph{(B)} and \emph{(D)} show $\xi$ \emph{vs.} $\langle |\Delta f| \rangle_{\phi}$. We quantify these relationships for all experimental packings using the Spearman's rank correlation, $\rho$. For the normal forces, the average correlations over packings for $\xi$ \emph{vs.} $\langle f \rangle_{\phi}$ and $\xi$ \emph{vs.} $\langle |\Delta f| \rangle_{\phi}$ are $\overline{ \rho_{f}} = 0.81$ and $\overline{ \rho_{\Delta f}} = 0.74$, respectively, with all $p$-values satisfying $p_{f} < 1 \times 10^{-174}$ and $p_{\Delta f} < 1 \times 1^{-127}$, respectively. (In Fig.~\ref{f:Spearman_flex} \emph{A,B} of the Appendix, we show the distributions of $\rho_{f}$ and $\rho_{\Delta f}$ for all packings). For the tangential forces, $\overline{ \rho_{f} } = 0.80$ and $\overline {\rho_{\Delta f} } = 0.71$, with all $p$-values satisfying $p_{f} < 1 \times 10^{-181}$ and $p_{\Delta f} < 1 \times 10^{-109}$. (In Fig.~\ref{f:Spearman_flex} \emph{C,D} of the Appendix we show the distributions of $\rho_{f}$ and $\rho_{\Delta f}$). In addition to the flexibility, we also tested the relationship between force and reconfiguration on a more robust measurement of local network rearrangement called \textit{promiscuity} \cite{promiscuity_code}, finding that the relationship still strongly holds. See Appendix~\ref{a:promiscuity} for a description of the promiscuity statistic, an example scatter plot, and correlation values.

To understand these results, first recall that the flexibility of a particle is a measure of how strongly fixed the particle is to its given community; $\xi$ is the number of times a particle changes community normalized by the number of possible changes, so lower values of $\xi$ correspond to particles that have more stable community allegiance across network layers. Our finding thus implies that large forces (or large changes in forces) are associated with particles reconfiguring and shifting communities throughout compression. 

\subsubsection{Mesoscale reconfiguration}

\begin{figure}
\includegraphics[width=\columnwidth]{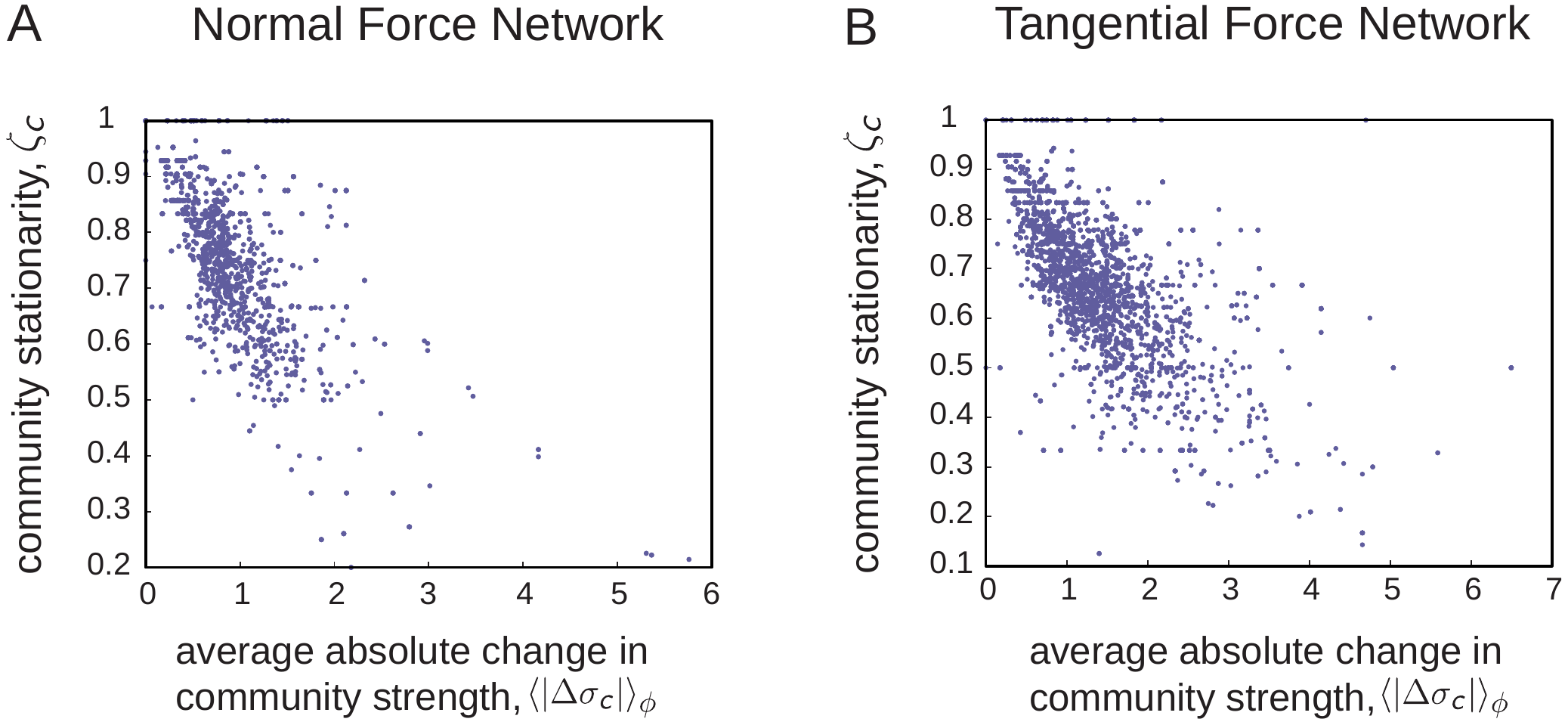}
\centering
\caption{\label{f:station_vs_diff_strength} \textbf{Community reorganization is driven by changes in intra-community force.} Scatter plots show the community stationarity, $\zeta_{c}$, \emph{vs.} the average absolute change in community strength across compression, $\langle |\Delta\sigma_{c}|\rangle_{\phi}$, for an example packing. For both the normal \emph{(A)} and tangential \emph{(B)} networks, there is a strong Spearman correlation between the two quantities (see Fig.~\ref{f:Spearman_diff_stats_all_life} for the distribution of correlations for each packing).}
\end{figure}

\begin{figure}
\includegraphics[width=\columnwidth]{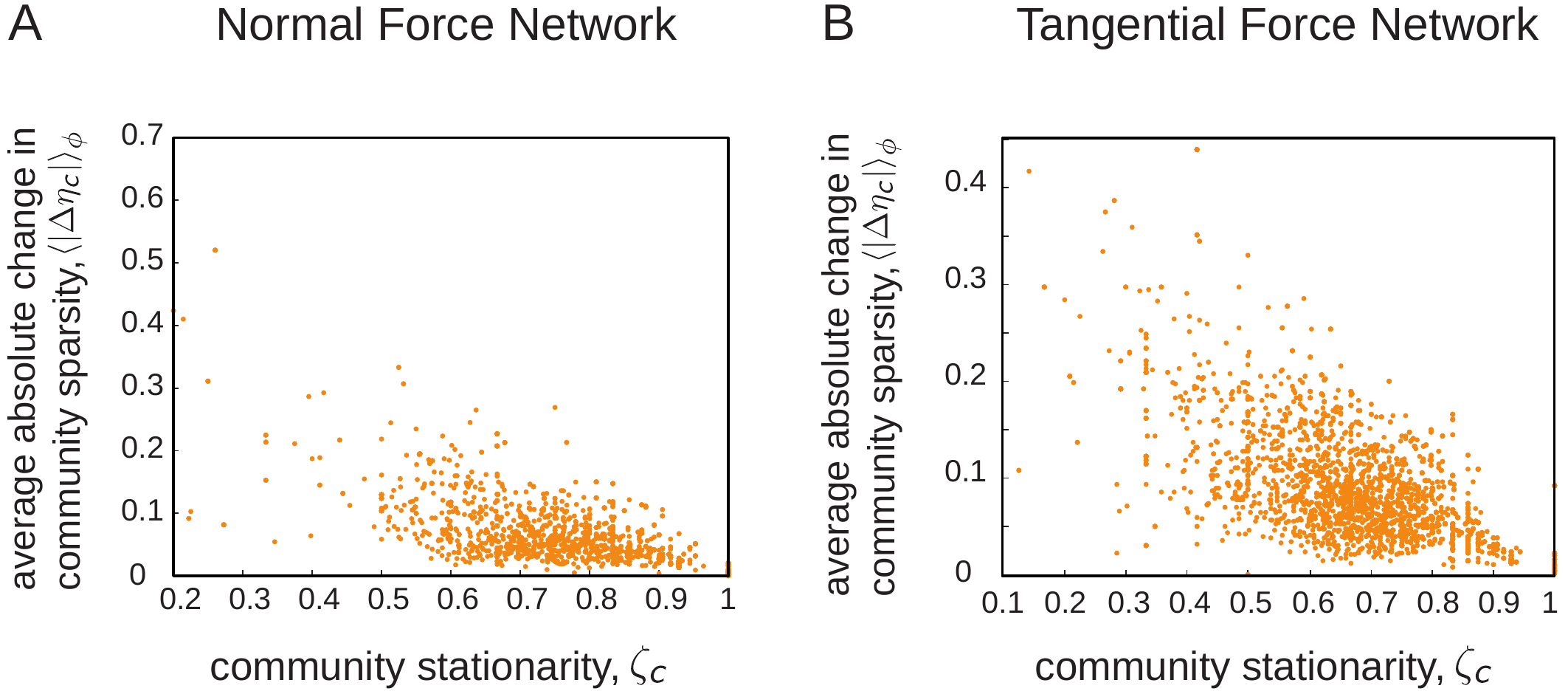}
\centering
\caption{\label{f:diff_sparsity_vs_station} \textbf{Changes in spatial arrangement across compression relate to community reorganization.} Scatter plots show the average absolute change in community sparsity across compression, $\langle |\Delta\eta_{c}|\rangle_{\phi}$, \emph{vs.} community stationarity, $\zeta_{c}$, for an example packing. For both the normal \emph{(A)} and tangential \emph{(B)} networks, there is a strong Spearman correlation between the two quantities. (See Fig.~\ref{f:Spearman_diff_stats_all_life} for the distribution of correlations for each packing).}
\end{figure}

We now examine the relationship between mesoscale reconfiguration as measured by the stationarity, and changes in physical properties of the multilayer community structure as measured by strength and sparsity. In particular, we first ask if reconfiguration at the community level is correlated with changes in the community strength throughout the compression process. We next investigate if changes in community particle composition drive changes in the spatial arrangement of community structure. For a given experimental run and modularity optimization, the stationarity ($\zeta$) for all communities is computed using Eq.\ref{eq: station}, and the strength ($\sigma$) and sparsity ($\eta$) of all communities at each layer are computed using Eqs. \ref{eq: comm_strength} and \ref{eq: comm_sparsity}, respectively. We then compute the Spearman correlation between stationarity, $\zeta_{c}$ and the average absolute change in strength $\langle |\Delta\sigma_{c}|\rangle_{\phi}$ and sparsity $\langle|\Delta\eta_{c}|\rangle_{\phi}$ across all layers in which a given community exists. The average absolute change in the strength of a community $c$ across applied strain steps is calculated according to
\begin{equation}
\langle |\Delta \sigma_{c}| \rangle_{\phi} = \frac {1} {L - 1} \sum_{l = 1}^{l = L - 1} |\sigma_{l+1}^{c} - \sigma_{l}^{c}|,
\label{eq: change_in_strength}
\end{equation}
and the average absolute change in sparsity is given by

\begin{equation}
\langle |\Delta \eta_{c}| \rangle_{\phi} = \frac {1} {L - 1} \sum_{l = 1}^{l = L - 1} |\eta_{l+1}^{c} - \eta_{l}^{c}|.
\label{eq: change_in_sparsity}
\end{equation}
Communities that have $s_{l}^{c} = 1$ for all layers are ignored.

We find that the stationarity is significantly anti-correlated with changes in community strength (a measure of intra-community force). Fig.~\ref{f:station_vs_diff_strength} shows scatter plots of $\zeta_{c}$ \emph{vs.} $\langle|\Delta\sigma_{c}|\rangle_{\phi}$ for all communities from each of the 20 optimizations of one experimental configuration for the normal force network \emph{(A)} and tangential force network \emph{(B)}. Specifically, large changes in $\sigma_{c}$ across compression give rise to lower values of $\zeta{c}$ (which corresponds to large changes in community particle content).  For the normal force network, the Spearman correlation averaged over optimizations and then packings for $\zeta_{c}$ \emph{vs.} $\langle|\Delta\sigma_{c}|\rangle_{\phi}$ is $\overline{\rho_{\Delta\sigma}} = -0.65$ with all $p$-values less than $0.0033$, and for the tangential network, $\overline {\rho_{\Delta\sigma}} = -0.69$ with all $p$-values less than $1 \times 10^{-13}$. Fig.~\ref{f:Spearman_diff_stats_all_life} \emph{A,C} in Appendix~\ref{a:Spearman_distributions} show the distributions of the optimization-averaged Spearman correlations for each experimental packing.

To examine the effect of mesoscale network reorganization on community spatial structure, we show an example of $\langle|\Delta\eta_{c}|\rangle_{\phi}$ \emph{vs.} $\zeta_{c}$ for one packing in Fig.~\ref{f:diff_sparsity_vs_station}.  Calculation of Spearman correlation values between the two statistics identifies a strong negative correlation between stationarity and sparsity, suggesting that changes in community particle content ($\zeta$) indeed lead to changes in spatial density ($\eta$). We find $\overline {\rho_{\Delta\eta}} = -0.77$ with all $p$-values less than $1 \times 10^{-10}$ for the normal force network, and $\overline {\rho_{\Delta\eta}} = -0.75$ with all $p$-values less than $1 \times 10^{-23}$ for the tangential forces. Fig.~\ref{f:Spearman_diff_stats_all_life} \emph{B,D} in Appendix~\ref{a:Spearman_distributions} show the distributions of the optimization-averaged Spearman correlations for each experimental packing.

The findings presented here demonstrate first that changes in community strength (i.e., in intra-community force), are associated with mesoscale reconfiguration across compression (i.e., the stationarity). This result complements those of Sec. \ref{s:local_reconfiguration}, where high values of force and change in force across compression were associated with high values of local reconfiguration -- a similar relationship holds for reconfiguration at the intermediate scale as well. We have additionally shown that if a community experiences large changes in its particle content, then the spatial arrangement of the community's particles undergoes restructuring as well, agreeing with physical intuition. It is thus clear that the reorganization of mesoscale network architecture is strongly tied to changes in physical properties of the mesoscale structure that occur due to the compression process.

\begin{figure}
\includegraphics[width=\columnwidth]{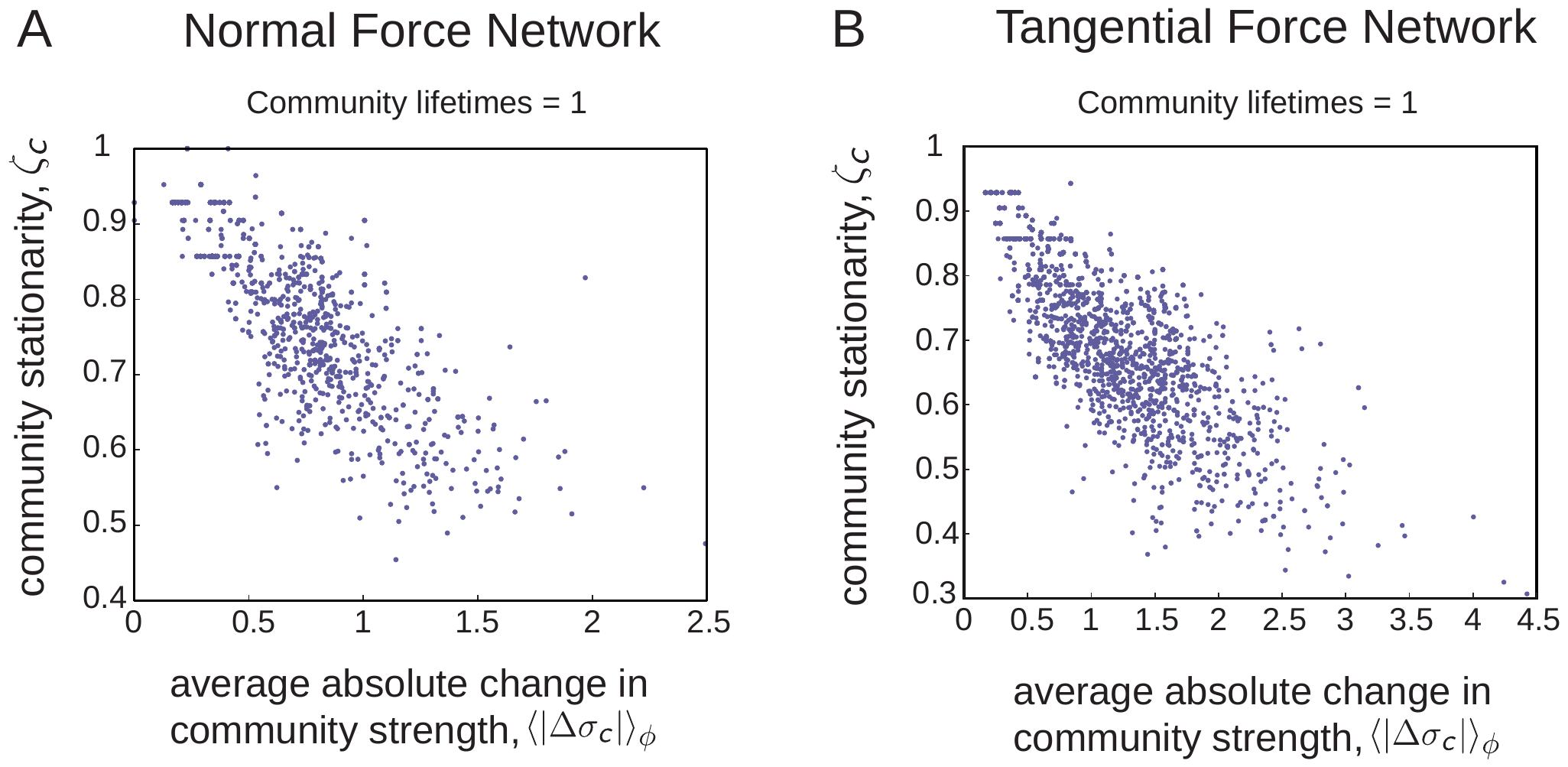}
\centering
\caption{\label{f:diff_stat_vs_station_life1} \textbf{Stationarity \emph{vs.} absolute change in community strength for long-lived communities.} Scatter plots show the community stationarity, $\zeta_{c}$, \emph{vs.} the average absolute change in community strength across compression, $\langle |\Delta\sigma_{c}|\rangle_{\phi}$, for communities that exist for the entire compression process. For both the normal \emph{(A)} and tangential \emph{(B)} networks, there is an increased correlation between the two quantities across packings.}
\end{figure}

It is important to point out that different communities might not exist for the same number of strain steps. Therefore, as a more robust measure of the relationship between community strength and stationarity, we repeat the above analysis but consider only those communities that persist throughout the entire compression cycle. If we define the \textit{lifetime} of a community to be the number of steps in which it exists divided by the total number of steps in the network, then we consider the set of communities that have lifetimes equal to 1. On average over all optimizations and experimental runs, these long-lived communities correspond to fractions of $\approx 0.62$ for the normal force network and $\approx 0.54$ for the tangential force network. For both the normal and tangential forces, we find that the correlations between community strength and stationarity increases compared to the correlation calculated using communities with all lifetimes. In particular, $\overline{\rho_{\Delta\sigma}} = -0.80$ with all $p$-values less than $1 \times 10^{-8}$ for the normal force network (see Fig.~\ref{f:diff_stat_vs_station_life1}\emph{A} for an example) and $\overline {\rho_{\Delta\sigma}} = -0.78$ with all $p$-values less than $1 \times 10^{-12}$ (see. Fig.~\ref{f:diff_stat_vs_station_life1}\emph{B} for an example) for the tangential force network. (Note that the horizontal ``stripes" at high values of $\zeta$ correspond mostly to communities with only a few particles; in the case of small groups, it is likely for different communities to achieve the same value of stationarity, because only so many reconfigurations are possible). The stronger correlations observed here may be due to the fact that they are inferred only from communities that exist for the same number of steps (and so are more comparable to one another), and also from the fact that these communities have the most data over which to compute averages. Physically, these longer-lived communities are also expected to undergo more continuous re-arrangements with compression, thus offering smooth estimates of stationarity; while shorter-lived communities may experience more extensive rearrangements, leading to their birth or death, and thus may offer more variable estimates of stationarity.  Fig.~\ref{f:Spearman_diff_strength_life1} in Appendix~\ref{a:Spearman_distributions} shows histograms of the Spearman correlations for all packings.

\subsection{Sensitivity to the subsystem}

A useful feature of our experimental setup is the presence of the low friction particles in the center of the packing. As a final demonstration of the sensitivity of the multilayer network model to the underlying physics, we examine its ability to detect differences in the physical properties of the subsystem (low $\mu$) \emph{vs.} the bath (high $\mu$). Sensitivity to such variation is crucial to the utility of this framework.

The first step is to consistently define the subsystem. The majority of subsystem particles are identified using UV dye, as described in Sec.~\ref{s:experimental_methods}. We use the UV dye image to initially identify particles as low friction or high friction particles at each strain step. Then, after having tracked particles in each frame, particles are labeled as low friction if they are identified as low friction for more than half of the steps. This technique ensures that low friction particles are not mislabeled as high friction particles, and the reverse rarely occurs. 

The second step is to define what constitutes a subsystem community \emph{vs.} a non-subsystem community. Since the communities we consider consist of more than one particle, it is highly unlikely for all particles within a group to either be only inside the subsystem radius or only outside the subsystem radius. We therefore consider a subsystem community, $c_{s}$, to be a community such that more than 75\% of its particles are contained within the low friction area at some point during the compression process. Any community $c_{b}$ that does not meet this condition is considered part of the bath. To ensure a fair comparison between these two groups, we further require that the communities examined in the two groups must be of similar sizes. (Due to the nature of the experimental setup, the subsystem communities are localized within a smaller region of space, and therefore will have a larger constraint on their possible sizes). Within each optimization, we compute the average size $\langle s_{c} \rangle_{\phi}$ of all communities across all steps in which the communities exist. We then find the largest and smallest average sizes of the subsystem communities, $\langle s_{c_{s}} \rangle_{\phi,max} $ and $\langle s_{c_{s}} \rangle _{\phi,min} $, and find all bath communities that have sizes $\langle s_{c_{b}} \rangle_{\phi}$ such that $\langle s_{c_{s}} \rangle_{\phi,min}  \ \leq \ \langle s_{c_{b}} \rangle_{\phi} \leq \ \langle s_{c_{s}} \rangle _{\phi,max} $. This protocol mitigates the likelihood that observed differences are due to differing community sizes.

\begin{figure}[!h]
\includegraphics[width=\columnwidth]{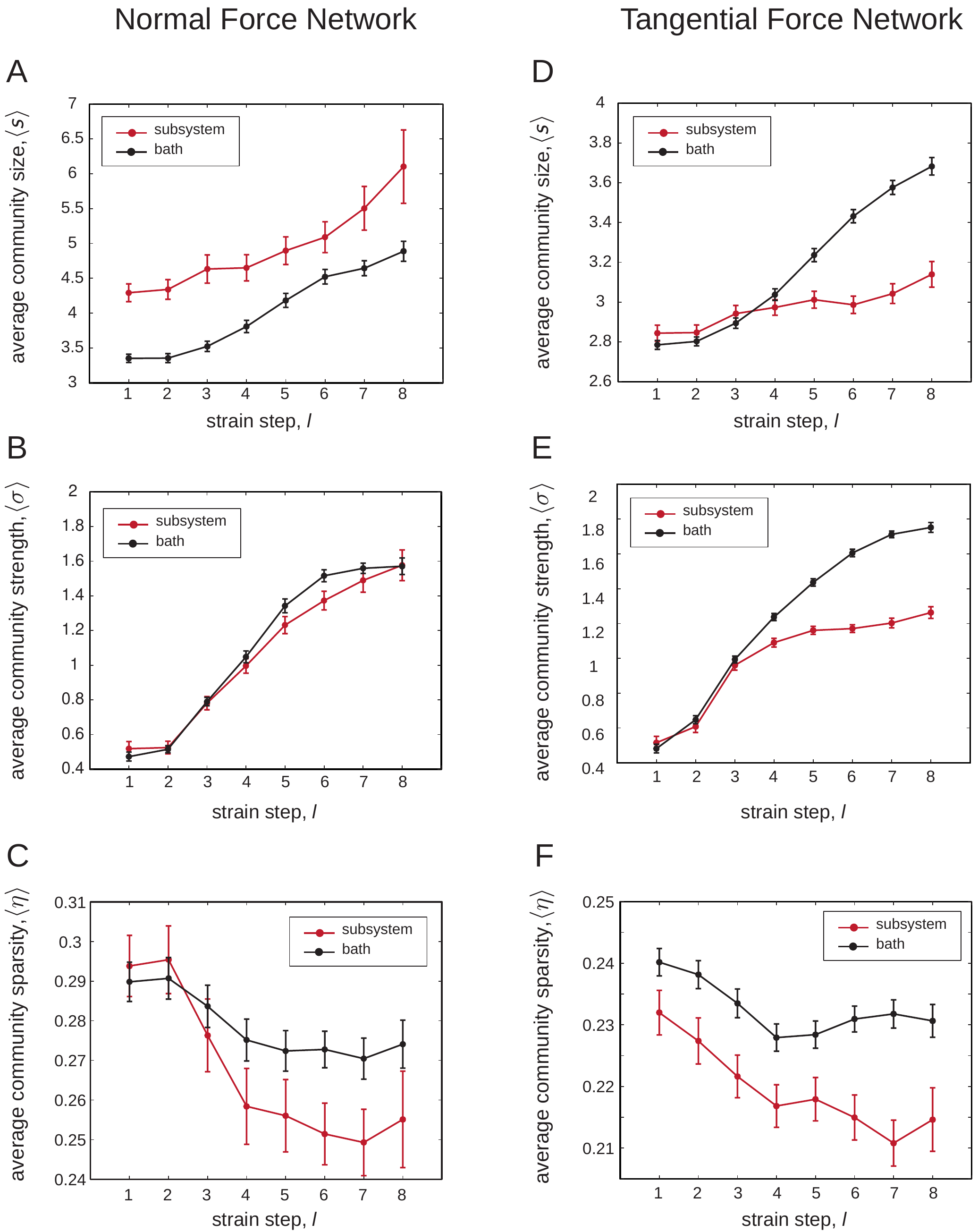}
\centering
\caption{\label{f:subsystem_comparison} \textbf{A comparison of the physical characteristics of the subsystem and bath communities}. \emph{(A, D)} Curves showing the average community size of the subsystem (red curve) \emph{vs.} bath (black curve) communities for the normal and tangential networks. The small differences between the curves for each network ensures that we compare communities of similar size between the two groups. \emph{(B, E)} The evolution of community strength in the network of normal forces is not significantly different between the subsystem and bath. However, the tangential network displays a clear separation, suggesting that the multilayer network model and community detection are sensitive to differences in friction. More importantly, the direction of the sensitivity agrees with physical intuition; the modular structures found in the low friction bath are characterized by lower tangential intracommunity force. \emph{(C, F)} There is a small distinction between the community sparsity of the two groups, with the low friction communities being slightly more dense throughout compression.}
\end{figure}

We now proceed to investigate if and how the community strength and sparsity evolve differently for the subsystem mesostructures \emph{vs.} those in the bath. We follow the procedure described in Sec. \ref{s:physical_properties}, but here we separate the subsystem and bath communities. In Fig.~\ref{f:subsystem_comparison}, we show the results of this analysis for both the normal \emph{(A-C)} and tangential \emph{(D-F)} force networks. For each of the three statistics previously considered, $\langle s \rangle$, $\langle \sigma \rangle$, and $\langle \eta \rangle$, we plot one curve for the bath (shown in black) and one for the subsystem (shown in red). Panels \emph{(A)} and \emph{(D)} show the evolution of community size; here we note the scale of the y-axes, which show that we are indeed comparing $c_{s}$ and $c_{b}$ of similar sizes (the curves remain within a size difference of about 1 particle throughout all steps).

We next examine the intra-community force of the high and low friction groups. Panels \emph{(B)} and \emph{(E)} show the average community strengths of the subsystem and bath. For the normal force networks, the two curves are very close; in other words, there does not appear to be a large difference in the evolution of the intra-community normal forces between the low and high friction mesostructures. This is not a surprising result, however, as we expect the difference in friction to more significantly effect the tangential component rather than the normal component.  Indeed, when we consider the network of tangential forces, we are clearly able to distinguish between the architecture of the subsystem and bath. At the beginning of the compression cycle, the average strength of the low and high friction communities still follow each other closely, but after the third step, we observe that the two groups separate. In addition to the community strengths being different, the \emph{way} in which they differ agrees with physical intuition; the intra-community tangential force of the high friction particle assemblies grows significantly more than the intra-community force of the low friction particle assemblies as compression continues. 

We also consider the average community sparsity in order to quantify differences in the spatial arrangement of the mesostructures. We observe that $\langle \eta \rangle$ tends to be smaller for the subsystem communities than those in the bath, although the differences are quite small. It is thus difficult to make a strong conclusion from the sparsity measure alone.

The findings discussed here first indicate that our model is indeed sensitive to underlying differences in the bath and the subsystem. But perhaps more crucial are the physical conclusions about network evolution; if only the network of normal forces is considered, it is difficult to discern the evolution of the low and high friction mesostructures. In this case, both groups display strong community organization throughout all applied strain steps. However, when the tangential force network is considered, the low friction subsystem forms noticeably weaker communities than the high friction bath, and the evolution of the two groups can be distinguished. This result suggests that the examination of \emph{both} the normal and tangential force networks can lead to a more complete understanding of mesoscale structure in granular systems. In Sec. \ref{s:discussion}, we discuss broader implications of the findings discussed in this section.

\section{Discussion and methodological considerations  \label{s:discussion}}

Although it is thought that mesoscale structures underlie many bulk properties of granular materials, there is a pressing need to develop theoretical tools for identifying these structures and describing their dynamics. Graph theoretic approaches provide a framework in which to address these challenges, via the network of contact forces connecting the grains. In this study, we use a common geometry (biaxial compression) as a case study for developing novel approaches drawn from network theory.

We describe the evolution of the granular system by modeling it as two temporal multilayer networks (one constructed from the normal component of the inter-particle forces, and the other from the tangential component). Then, we extract inherent mesoscale structure from each network using multilayer community detection techniques with a physically grounded null model. Unlike in a static network, the particle assemblies we detect persist across layers, allowing a direct study of the evolution of mesoscale organization. The particle assemblies are quantified with measures of network reconfiguration, as well as with measures of physical architecture. We find that both the normal and tangential components display community structure, though the network of normal forces shows organization on a larger scale (bigger communities), and exhibits less restructuring during the sequence of applied strain steps. We also demonstrate that the community architecture of the real system is significantly distinguishable from three physically-relevant null models.  

In addition to characterizing the evolution of the granular network in terms of force and the spatial arrangement of particle assemblies, we show that network reconfiguration can be tied to these physical descriptors. This is a crucial finding, as it bridges the gap between seemingly abstract measures of network reorganization and physical properties of network components. One result in particular that merits discussion is that particle flexibility is positively correlated with the average force on a particle across layers.  Although the notion that force drives local reorganization is intuitive from a physical standpoint, it is interesting to consider this result in the context of a different type of network where flexibility has also been studied. Since we consider particles to be nodes and forces to be weighted edges, there is a direct analogy between the total force on particle $i$ in a given static network, and the \textit{weighted degree} of node $i$, $k_{i} = \sum_{j} A_{ij}$ in a general undirected network with adjacency matrix $\mathbf{A}$. In a study on the dynamic community structure of functional human brain networks during learning, a seemingly opposite relationship between flexibility and node strength exists \cite{Bassett:2013ba}. It was found that nodes from densely connected brain regions exhibited little change in community allegiance throughout learning, whereas weakly connected nodes had higher values of flexibility. This points to an interesting contrast that exists between the evolution of the granular network and the much less physically constrained functional brain network. It is reasonable to think that the source of the opposing relationship between node strength and flexibility is due to the physical nature of the granular system, and the interpretation of node strength as force, which drives reconfiguration. But perhaps more importantly, this finding motivates the general need to develop more physically informed network measures to study these mechanical systems. The use of the geographic null model (vs. the traditional Newman-Girvan model) is an important step.

In the final section, we consider the presence of the low friction subsystem, asking \textit{if} our model is sensitive to the subsystem, and more importantly, \textit{how}. We find that from the normal force network alone, it is difficult to distinguish the two groups. However, consideration of the tangential network shows a clear difference in the evolution of the intra-community forces of the subsystem versus the bath, with the low friction modules characterized by noticeably smaller tangential strength. These conclusions illustrate that low-friction systems can have different community organization, and that the both the normal and tangential networks contain useful information. An interesting direction for future work would be to examine the evolution of sheared systems using the multilayer network approach, where we might expect to see heightened sensitivity to differences in friction, and perhaps altered reconfiguration patterns.

It is our hope that the method and results presented in this work will inform future studies on granular materials. For example, an interesting question is whether or not measures of network reconfiguration can be related to other notions of failure and deformation, like soft spots \cite{Manning:2011aa} and force chain buckling \cite{Tordesillas:2007aa,Tordesillas:2009aa}. Better yet, can we \textit{predict} these rearrangement events from structural features of the network before deformation? Investigation of the latter will likely require the formulation of additional physically informed statistics of network dynamics, and may moreover be complemented by machine learning techniques. The fact that the multilayer network framework is sensitive to differences in the reconfiguration of null models as well as to differences in friction, further suggests that the framework may find applicability beyond this study. One can imagine using this formulation to not only characterize the evolution of a single system, but also to distinguish, quantify, and classify dissimilarities between several systems that are slightly different. For example, compressed systems \emph{vs.} sheared systems, or experiments \emph{vs.} simulations. One could also use this framework to examine how varying particle shapes and sizes effect the evolution and structure of the force network at the intermediate scale, or to inform the design of materials that exhibit tailored physical properties under external perturbations. Finally, the ability to identify and quantify mesoscale architecture is important in other physical systems more broadly. The brain \cite{Sporns:2013aa, Sporns:2016aa, Bassett:2015ab, Bassett:2011aa, Bassett:2013aa} and biomaterials \cite{Pong:2011, Quinn:2011, Han:2013, Zhao:2014, Zhang:2015} are a few examples of systems where intermediate structure is crucial to function.

In order to solidify the utility of the model presented here, it will be necessary to carry out a deeper exploration of the methods. The investigation of different null models is one place to start, especially in regard to the tangential network. In this study we took the absolute value of the tangential forces in order to use the geographic null model and directly compare to the normal forces, but in different situations (sheared systems, for example) the \textit{sign} of the force is important and should not be discarded. However, to keep this information will require a null model that can correctly deal with signed adjacency matrices without losing physical meaning. A more thorough sampling of the phase space formed by the two resolution parameters, $\omega$ and $\gamma$, would also be useful. It will be interesting to consider the robustness and clarity of results with respect to changes in these quantities. Additionally, it may be more physically reasonable to move away from a scalar temporal coupling and examine more complex (but perhaps more informative) ways of linking layers together. For example, particle properties or similarity measures between the individual static networks could be used to determine the magnitude of coupling strengths in the multilayer formulation. Finally, it will be important to continue developing relevant measures of multilayer network reconfiguration that are informed by the physical questions we are attempting to answer.

In this study, we have utilized multilayer networks to model and characterize the structural reconfiguration that occurs in a compressed granular material. The framework introduced here is sensitive to the important inhomogeneity present in granular media, and allows for the direct extraction of evolving, intermediate scale structure that has not previously been probed or examined. We conclude by noting that in addition to particulate systems, the methods developed and exercised in this work can be applied more broadly across physical, biological, technological, and social systems to understand how dynamic interactions between system components organize into collective structure, which in turn constrains bulk properties and system performance.

\section*{Acknowledgements}

We thank Eli T. Owens, Chad Giusti, and Shi Gu for helpful conversations. LP and DSB acknowledge support from the John D. and Catherine T. MacArthur Foundation, the Alfred P. Sloan Foundation, the Army Research Laboratory and the Army Research Office through contract numbers W911NF-10-2-0022 and W911NF-14-1-0679, the National Institute of Mental Health (2-R01-DC-009209-11), the National Institute of Child Health and Human Development (1R01HD086888-01), the Office of Naval Research, and the National Science Foundation (PHY-1554488, BCS-1441502, BCS-1430087). KD is grateful for support from the National Science Foundation (DMR-0644743, DMR-1206808) and the James S. McDonnell Foundation. 


%


\clearpage
\appendix

\section{Additional figures of community structure}

\begin{figure*}
\includegraphics[width=\textwidth]{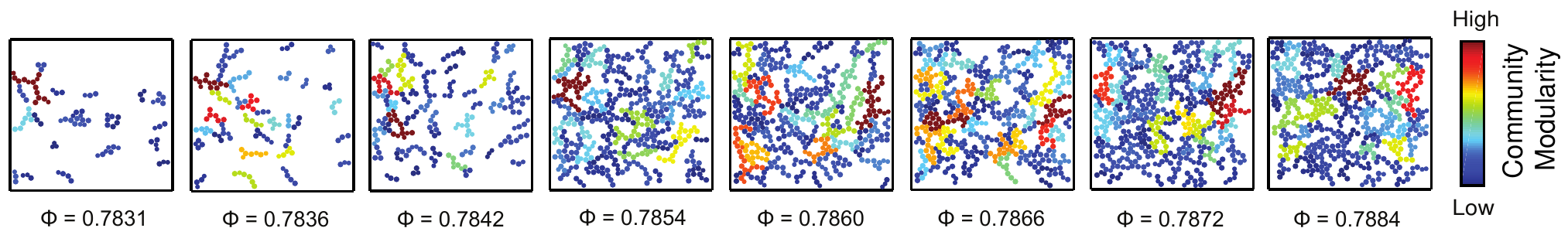}
\centering
\caption{\label{f:static_tan_comms} \textbf{An example of static community structure from the tangential force network.} At $\omega = 0$ we recover the results of single layer community detection; there is no notion of linking mesostructures from one compressive step to the next. }
\end{figure*}

\begin{figure*}
\includegraphics[width=\textwidth]{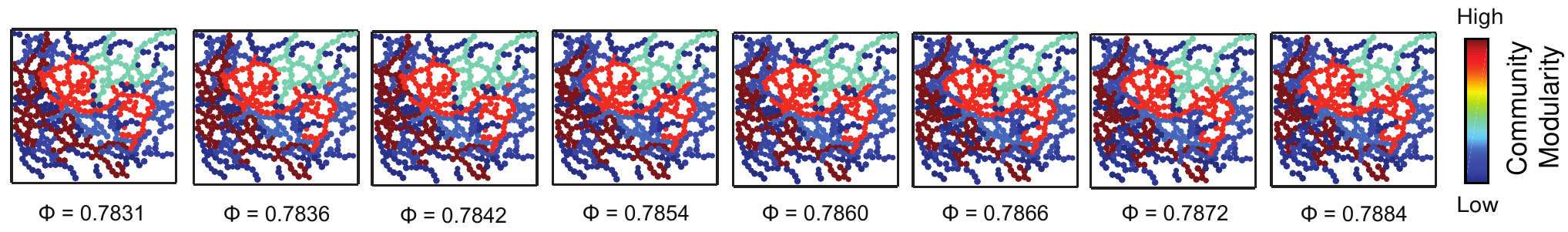}
\centering
\caption{\label{f:multilayer_comms_constant} \textbf{An example of the multilayer community structure of the normal force network at high inter-layer coupling ($\omega = 1$)}. At large enough values of $\omega$, the community structure becomes temporally consistent (unchanging across network layers), and the model becomes insensitive to changes in network architecture.}
\label{f:omega_1}
\end{figure*}

On the following page, we provide additional examples of community structure. Fig.~\ref{f:static_tan_comms} shows the results of community detection on the tangential force networks when the inter-layer coupling is zero; in this case, we recover the results of single layer community detection. As was the case with the normal force communities at $\omega = 0$ (Fig.~\ref{f:static_norm_comms}), the communities at each step are completely independent of one another. This fact motivates the generalization of the network framework to the multilayer regime, which allows for direct observation of evolving mesostructures throughout the compression cycle. Fig.~\ref{f:omega_1} shows an example of community structure detected at a high inter-layer coupling ($\omega = 1$), for the network of normal forces. In this case, the particle assemblies are almost completely consistent across layers, and we thus do not probe network reconfiguration.

\clearpage

\section{Effect of small omega on network flexibility}
\label{a:choosing_omega}

In Sec.\ref{s:choosing_omega} of the main text, we began the sweep over inter-layer couplings at $\omega = 0.01$, which corresponds to a weight equivalent to $\frac{1}{100}$ of the mean edge weight in each layer. Although this value is a relatively small step away from $\omega = 0$ (where the average network flexibility is $\overline{\Xi} = 1$), we find that even at such a small coupling, the flexibility drops significantly, to $\overline{\Xi} \approx 0.3349$. Here, we consider whether or not the flexibility can be brought closer to unity by bringing the coupling arbitrarily close to zero. 

\begin{figure}[!h]
\includegraphics[width=0.6\columnwidth]{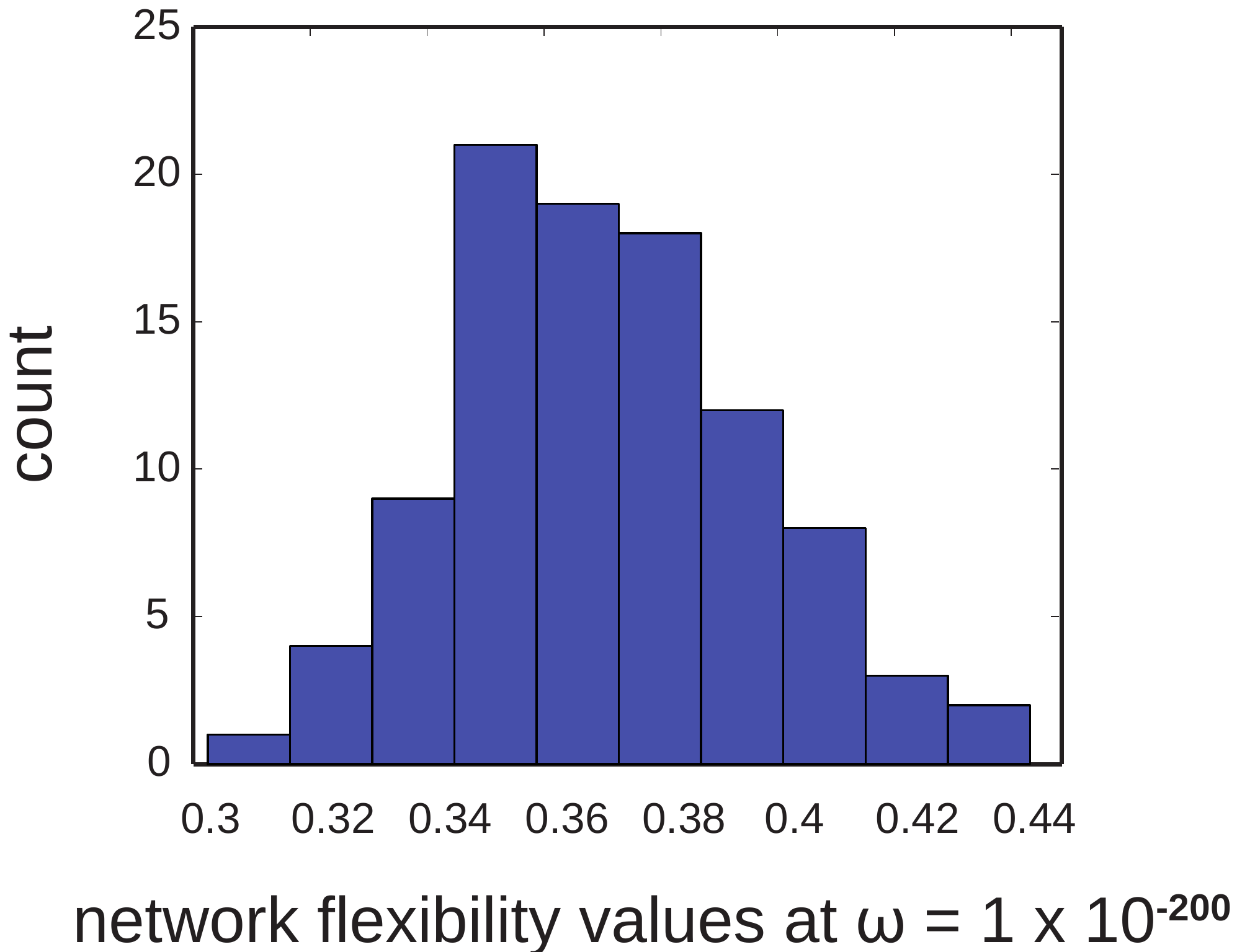}
\centering
\caption{\label{f:flex_small_omega} \textbf{Histogram of flexibility values over experimental packings for $\mathbf{\omega = 1 \times 10^{-200}}$}. Even at a very small inter-layer coupling, the flexibility values of the granular networks do not increase by a significant amount compared to the values obtained at $\omega = 0.01$.}
\end{figure}

To test the nature of network dynamics at small coupling, we carry out 20 optimizations of modularity maximization for each packing at $\omega_{\epsilon} = 1 \times 10^{-200}$ (using the network of normal forces). The flexibility of each optimization is then calculated according to Eq.~\ref{eq: flex}, and a single value for each configuration is obtained by averaging the values over the 20 realizations of community detection. Finally, we compute the average flexibility values over packings to obtain $\overline{ \Xi}$. Interestingly, we find that in the case of these granular networks, the mean flexibility over experimental configurations still remains much less than one, with $\overline{\Xi(\omega_{\epsilon})} = 0.37$. Fig.~\ref{f:flex_small_omega} shows the distribution of the optimization-averaged flexibility values, $\langle \Xi \rangle _{opt}$ for each packing. This finding suggests that even arbitrarily small values of the inter-layer coupling do not significantly increase the amount of reconfiguration observed in the resulting community structure, and solidifies the validity of our choice of initial $\omega$ at $0.01$.

\section{Evolution of physical properties and network dynamics}

\subsection{Communities with linear trends in physical characteristics}
\label{a:table_trend}

In Sec.~\ref{s:phys_trends} of the main text, we examine communities that exhibit linear trends with respect to strength, $\sigma$ and sparsity $\eta$, finding that communities with decreasing strength very often also have decreasing sparsity, and about half of communities with increasing sparsity also have increasing strength. There are many other possible combinations one could consider; here, we report percentages that quantify the likelihood that any two such combinations of linear trends occur together in a given community (\ref{t:trend_together}). To understand the following table, first recall that the quantity $\sigma_{\uparrow} \cap \eta_{\uparrow}$, for example, is the number of communities with increasing strength and increasing sparsity. Then, the quantity $(\sigma_{\uparrow} \cap \eta_{\uparrow})/\sigma_{\uparrow}$ gives the fraction of all communities with linearly increasing strength that also have linearly increasing sparsity. 

\begin{table}[!h]
\begin{centering}
\begin{tabular}{p{3cm} p{3cm}  l}
\hline \hline
Trend & Normal& Tangential \\
\hline
$\sigma_{\uparrow} \cap \eta_{\uparrow}/\sigma_{\uparrow}$  &  23.5  $\pm$ 0.6 & 41.2 $\pm$ 1.0 \\
$\sigma_{\uparrow} \cap \eta_{\downarrow}/\sigma_{\uparrow}$ & 46.2 $\pm$ 0.7 & 21.8 $\pm$ 0.5 \\
$\sigma_{\downarrow} \cap \eta_{\uparrow}/\sigma_{\downarrow}$ & 1.9 $\pm$ 0.4 & 2.3 $\pm$ 0.6 \\
$\sigma_{\downarrow} \cap \eta_{\downarrow}/\sigma_{\downarrow}$ & 83.3 $\pm$ 2.2 & 76.3 $\pm$ 2.1 \\
\\
$\eta_{\uparrow} \cap \sigma_{\uparrow}/\eta_{\uparrow}$  &  63.8 $\pm$ 1.2 & 50.4 $\pm$ 0.9 \\
$\eta_{\uparrow} \cap \sigma_{\downarrow}/\eta_{\uparrow}$ & 0.6 $\pm$ 0.1 & 0.3 $\pm$ 0.1 \\
$\eta_{\downarrow} \cap \sigma_{\uparrow}/\eta_{\downarrow}$ & 36.3 $\pm$ 0.7 & 14.3 $\pm$ 0.3 \\
$\eta_{\downarrow} \cap \sigma_{\downarrow}/\eta_{\downarrow}$ & 6.4 $\pm$ 0.5 & 5.5 $\pm$ 0.3 \\
\hline \hline
\end{tabular}
\caption{\label{t:trend_together} The percentage of communities that progress together in terms of their linear trends with respect to strength $\sigma$ and sparsity $\eta$. The first column denotes the relationship that we consider. The second and third columns are the average percentages over optimizations and packings for the given relationship, for the normal and tangential networks, respectively. Errors are the standard errors of the mean over packings.}
\end{centering}
\end{table}

\subsection{Promiscuity as a more robust measure of local reconfiguration}
\label{a:promiscuity}

Throughout this work, we use the notion of flexibility (Sec.~\ref{s:choosing_omega}) as a measure of local reconfiguration that is determined from the underlying mesoscale community structure in the system. As a more robust measure of local network reorganization, we also consider the \textit{particle promiscuity}. The promiscuity $\psi_{i}$ of particle $i$ is defined as the fraction of all communities in which the particle participates at least once, across all network layers. In the context of the current study, the promiscuity clarifies whether a particle is simply bouncing back and forth between the same two communities (which would yield high $\xi$ but low $\psi$), or truly participating in many different communities throughout compression (a situation that leads to both high $\xi$ and high $\psi$). As with the flexibility, we define the network promiscuity $\Psi$ to be the average over all particles,

\begin{equation}
\Psi = \frac{1}{N}\sum_{i}{\psi_{i}}.
\label{eq:prom}
\end{equation}

\begin{figure}[!h]
\centering
\includegraphics[width=\columnwidth]{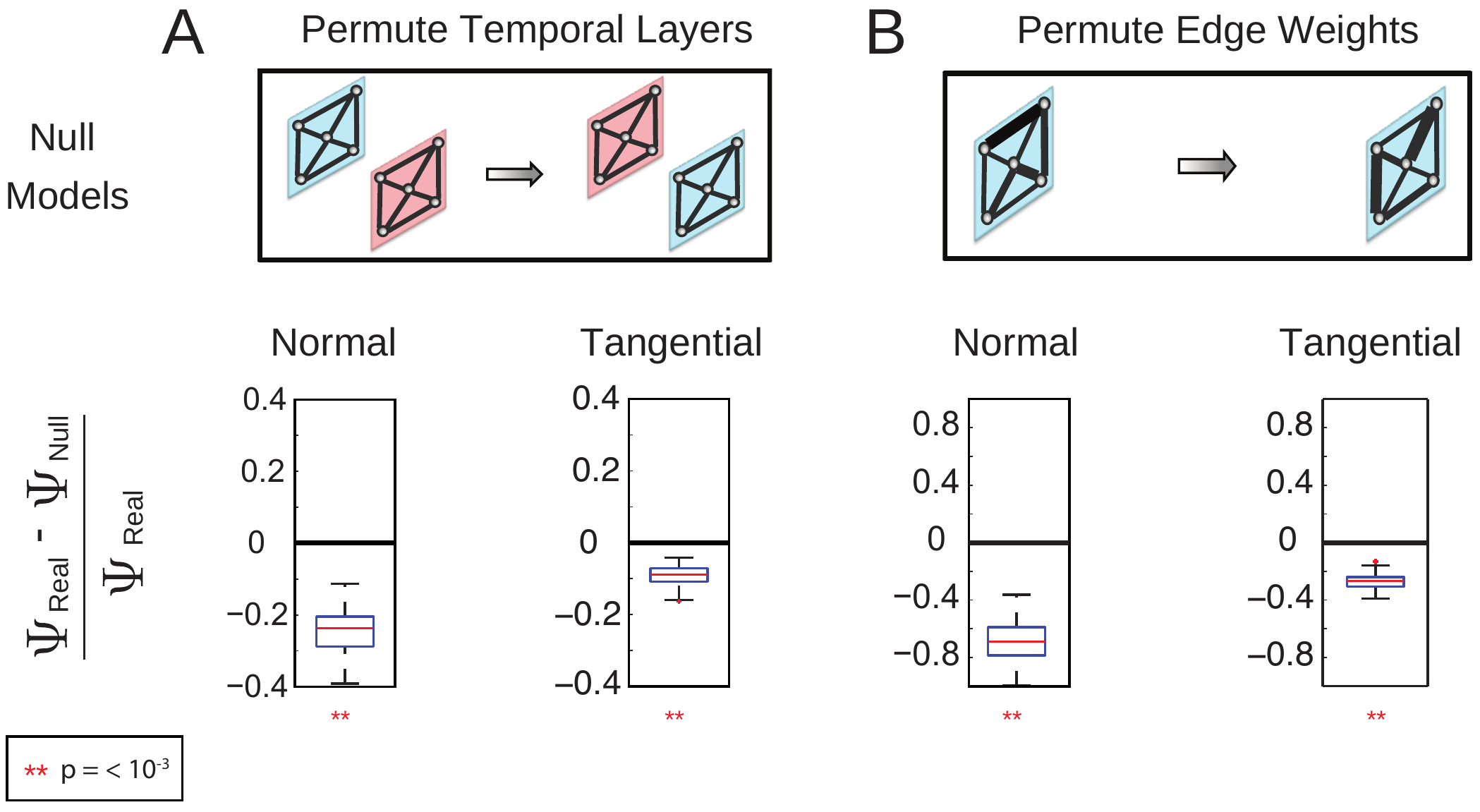}
\caption{\label{f:prom_null} \textbf{Network promiscuity values differ between the granular system and null models.} Significant differences between the reconfiguration of the granular network and the null models exist when the boxplots do not cross the zero line.  \emph{(A)} Boxplots showing that $\Psi_{real} < \Psi_{null}$ for the normal and tangential networks, where the null model, $\mathcal{A}^{temporal}$, is constructed by scrambling network layers. \emph{(B)} Boxplots showing that $\Psi_{real} < \Psi_{null}$ for the normal and tangential networks, where the null model, $\mathcal{A}^{edges}$, is constructed by scrambling the edge weights within each layer while preserving contact structure. In all cases, statistical significance is determined by permutation testing.}
\end{figure}

We first test whether the promiscuity differs between the multilayer force networks of the real system, and the temporal ($\mathcal{A}^{temporal}$) and force-scrambled ($\mathcal{A}^{edges}$) null models. In both cases, we follow the analysis described in Sec.~\ref{s:null_models}, and obtain results in line with those found using the flexibility. In particular, for both the normal and tangential force networks, the promiscuity values of the real system are lower than the promiscuities of both null models. In Fig.~\ref{f:prom_null} we show boxplots over the experimental configurations of $(\Psi_{real} - \Psi_{null})/\Psi_{real}$. (For each configuration, the value we consider is the average over community detection optimizations). The promiscuity values of the real and null networks are significantly different when the boxplots do not cross the zero-line.

\begin{figure}[!h]
\centering
\includegraphics[width=\columnwidth]{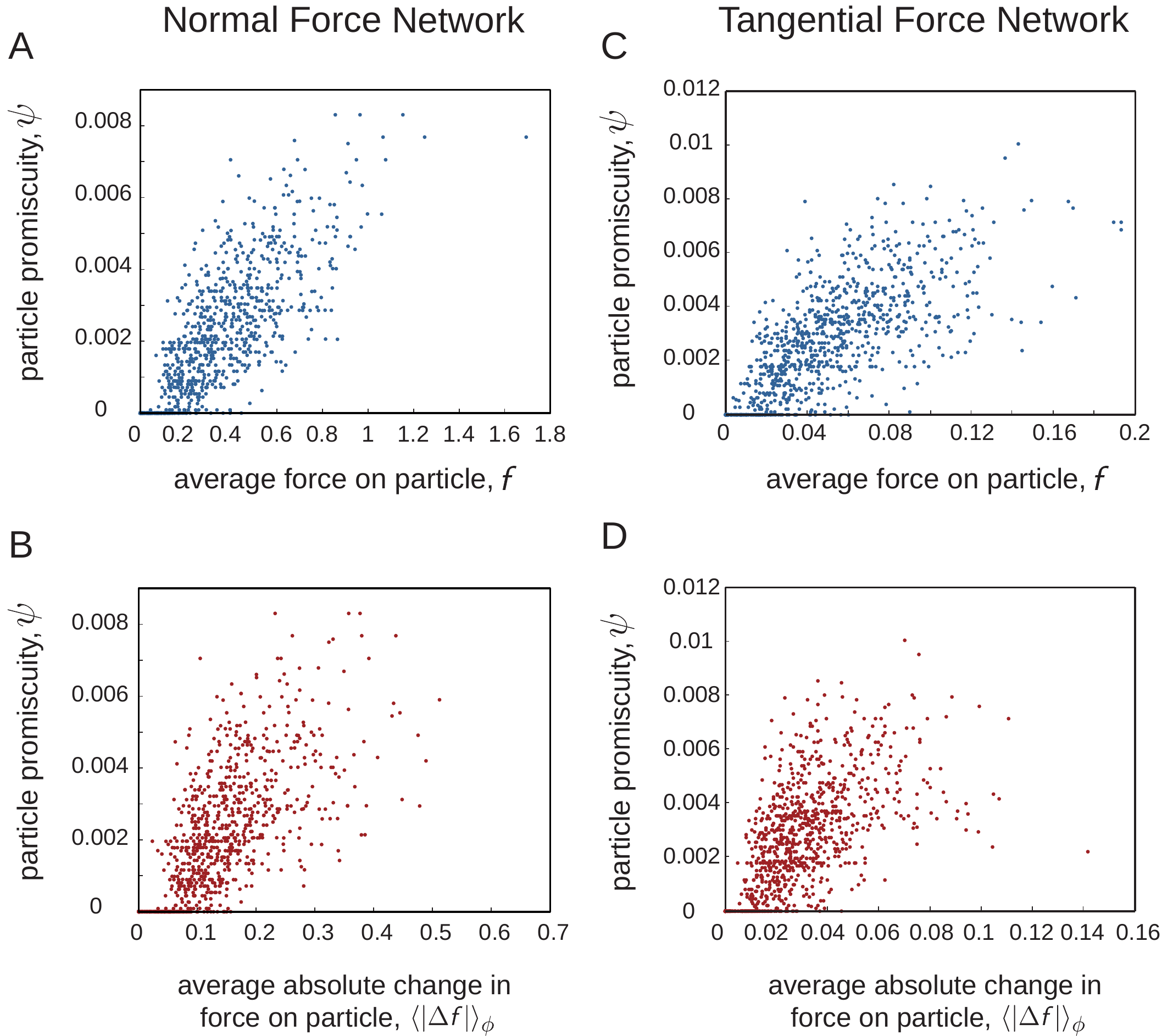}
\caption{\label{f:prom_vs_force} \textbf{Inter-particle forces predict a more robust measure of local reconfiguration.} The relationship between local network reconfiguration and the average force on a particle across network layers holds for a more robust measure of reorganization known as promiscuity. \emph{(A, C)} Scatter plots of particle promiscuity $\psi$ vs. the average force on a particle across all compressive steps $\langle f \rangle_{\phi}$ for a sample packing. For both the normal and tangential networks, there is a strong Spearman correlation between the two quantities. \emph{(B, D)} Scatter plots of particle promiscuity $\psi$ vs. the average absolute change in force on a particle across compression $\langle | \Delta f | \rangle_{\phi}$, for a sample packing. For both the normal and tangential networks, there is a strong Spearman correlation between the two quantities.}
\end{figure}

We also investigated whether or not the strong positive correlation between inter-particle force and flexibility also holds for inter-particle force and promiscuity. Carrying out the same analysis as described in Sec. \ref{s:local_reconfiguration}, we find that the result is robust: high values of the average force or absolute change in force across compression are positively correlated with high values of $\psi$. In particular, we find $\overline{ \rho_{f}} = 0.81$ and $\overline{ \rho_{\Delta f}} = 0.74$, with all $p$-values satisfying $p_{f} < 1 \times 10^{-173}$ and $p_{\Delta f} < 1 \times 10^{-128}$ for the network of normal forces, and $\overline{ \rho_{\Delta f}} = 0.79$ and $\overline{\rho_{\Delta f}} = 0.70$, with all $p$-values satisfying $p_{f} < 1 \times 10^{-169}$ and $p_{\Delta f} < 1 \times 10^{-107}$ for the network of tangential forces. We show an example scatter plot of these relationships for the particles in a single experimental packing in Fig.~\ref{f:prom_vs_force}, and show the distribution of Spearman correlations for each packing in Fig.~\ref{f:Spearman_prom}.

\subsection{Spearman correlation distributions}
\label{a:Spearman_distributions}

\begin{figure}[!h]
\centering
\includegraphics[width=\columnwidth]{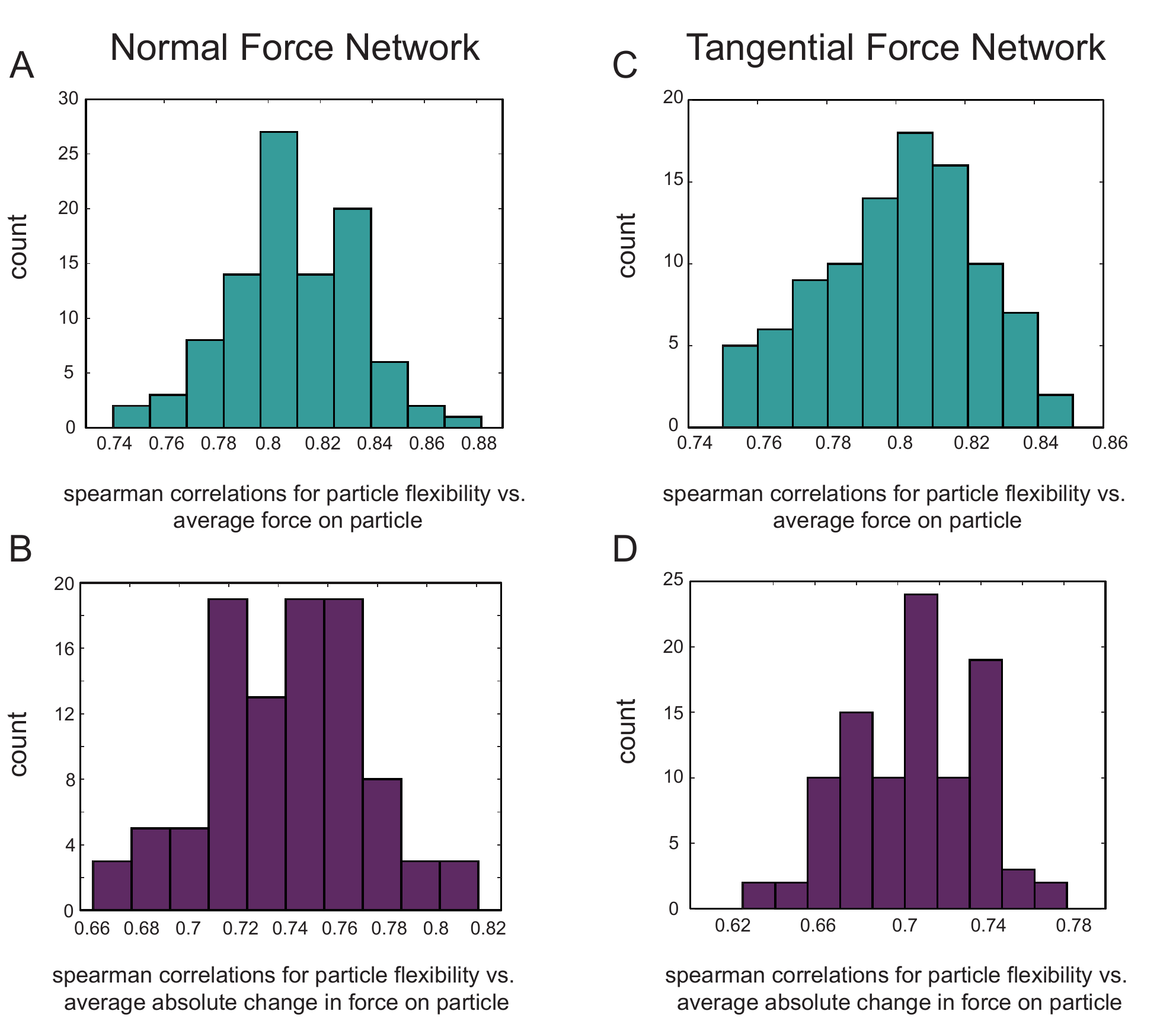}
\caption{\label{f:Spearman_flex} \textbf{Distribution of Spearman correlations for $\xi$ vs. $\langle f \rangle_{\phi}$ and $\langle | \Delta f | \rangle _{\phi}$ for all experimental packings}. \emph{(A)} Correlations for the normal force network. \emph{(B)} Correlations for the tangential force network. \\ \\}
\end{figure}

\begin{figure}[H]
\centering
\includegraphics[width=\columnwidth]{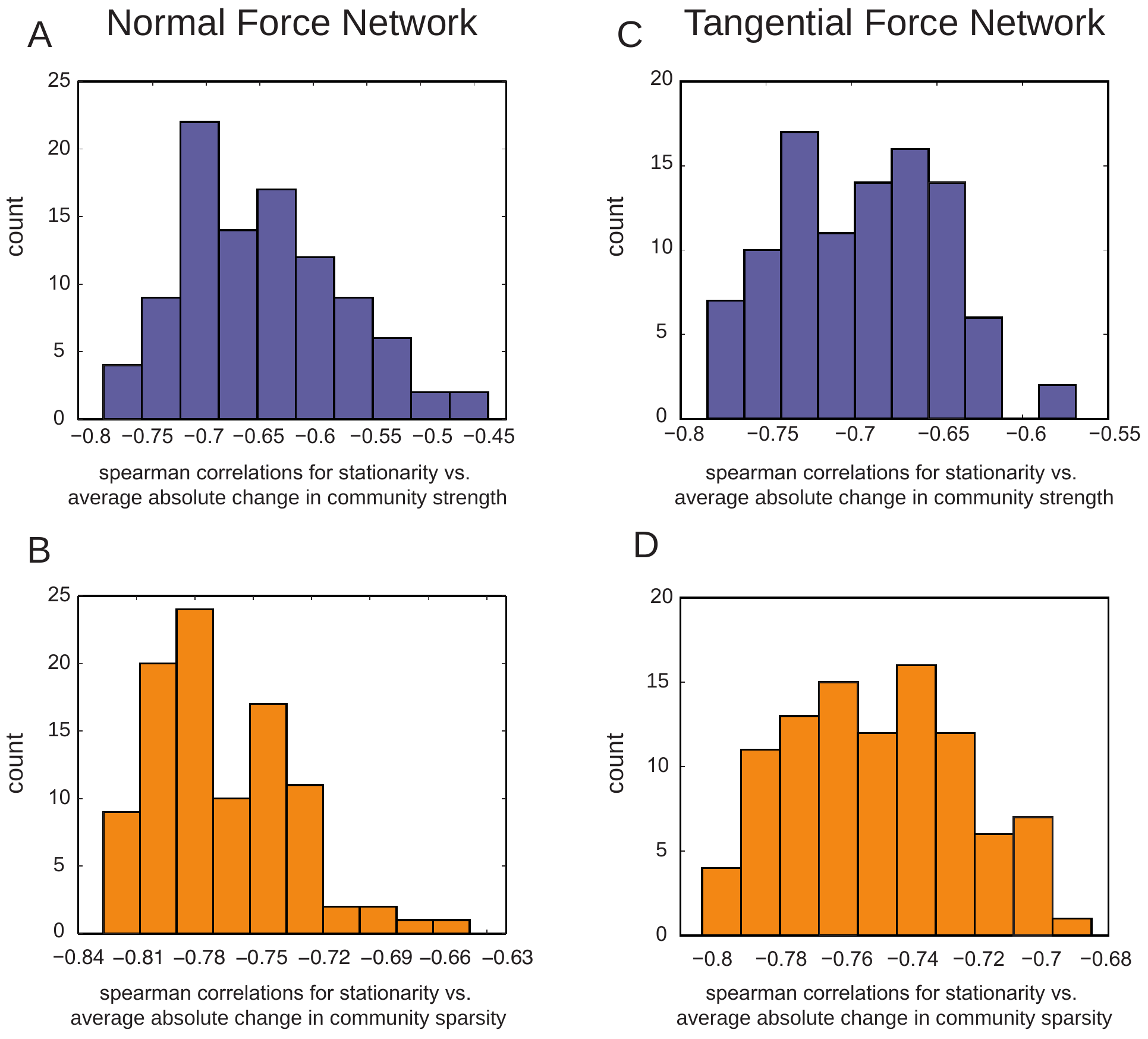}
\caption{\label{f:Spearman_diff_stats_all_life} \textbf{Distribution of Spearman correlations for $\zeta_{c}$ vs. $\langle |\Delta \sigma_{c}| \rangle _{\phi}$ and $\langle |\Delta \eta_{c}| \rangle _{\phi}$ vs. $\zeta_{c}$ when we include communities with all lifetimes}. The Spearman coefficients for the normal force network \emph{(A, B)} and tangential force network \emph{(C, D)} have been averaged over optimization for each packing.}
\end{figure}

\begin{figure}[!h]
\centering
\includegraphics[width=\columnwidth]{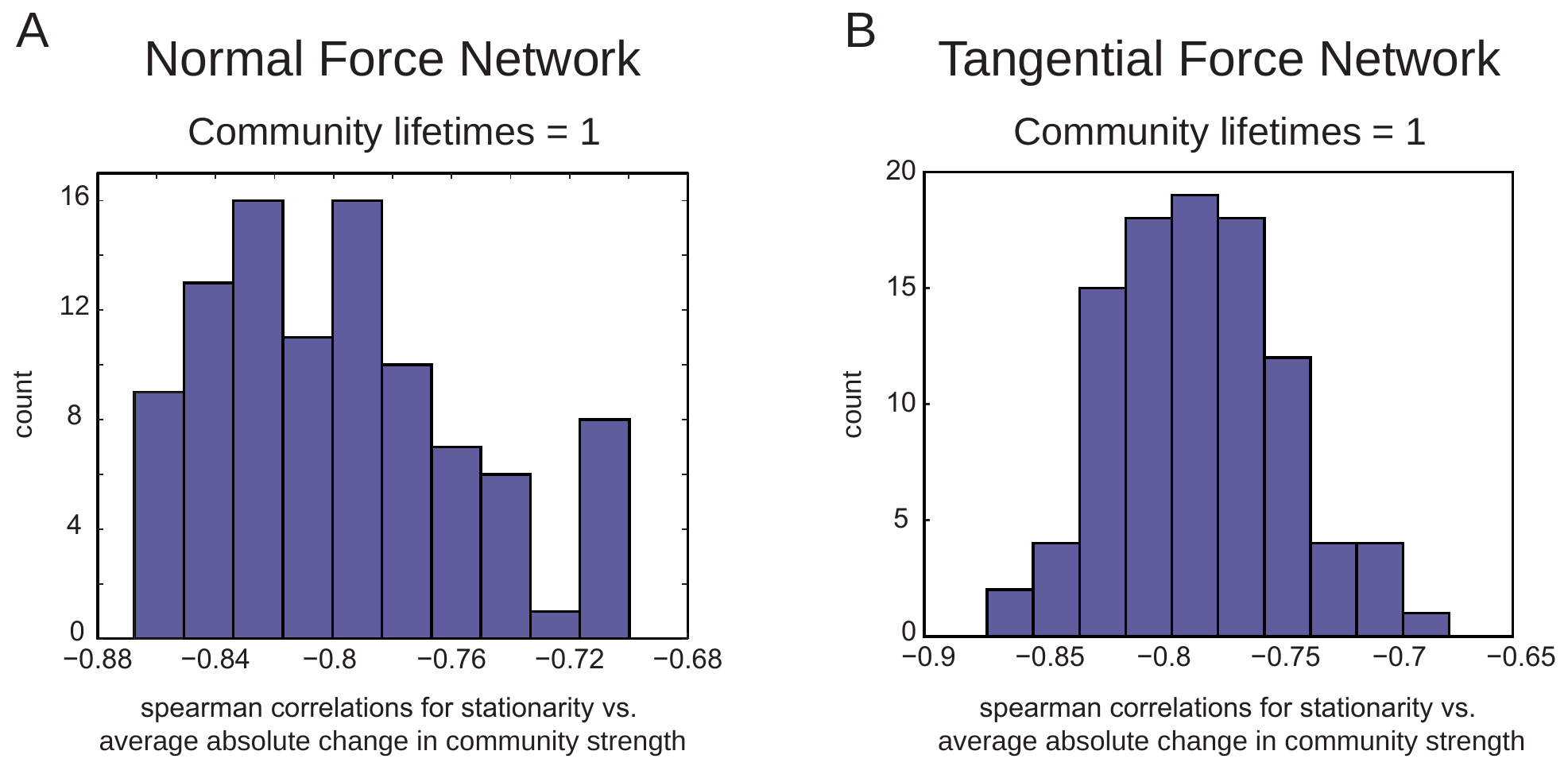}
\caption{\label{f:Spearman_diff_strength_life1} \textbf{Distribution of Spearman correlations for $\zeta_{c}$ vs. $\langle |\Delta \sigma_{c}| \rangle _{\phi}$ and $\langle |\Delta \eta_{c}| \rangle _{\phi}$ vs. $\zeta_{c}$ when we only include communities with lifetimes = 1}. The Spearman coefficients for the normal force network \emph{(A, B)} and tangential force network \emph{(C, D)} have been averaged over optimization for each packing.\\ \\}
\end{figure}

\begin{figure}[H]
\centering
\includegraphics[width=\columnwidth]{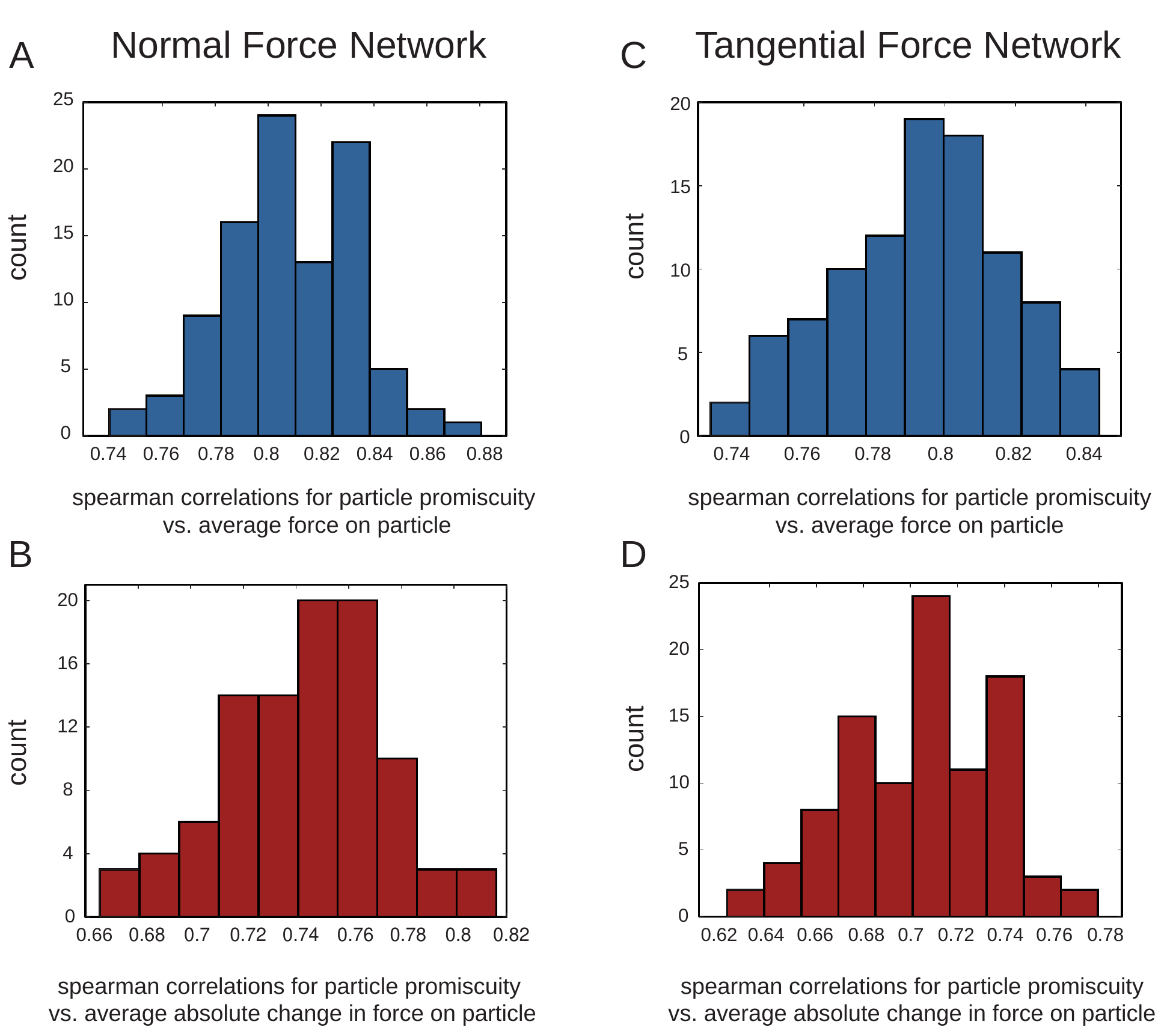}
\caption{\label{f:Spearman_prom} \textbf{Distribution of Spearman correlations for $\psi$ vs. $\langle f \rangle_{\phi}$ and $\langle | \Delta f | \rangle _{\phi}$ for all experimental packings}. \emph{(A), (C)} Correlations for the normal force network. \emph{(B), (D)} Correlations for the tangential force network.}
\end{figure}

Throughout this work, we quantify relationships between network reconfiguration and physical properties of network structure using Spearman correlations and example scatter plots. In the following figures, we show histograms of the Spearman correlation values for these relationships for all experimental packings. In particular, we show flexibility \emph{vs.} force (Fig.~\ref{f:Spearman_flex}), stationarity \emph{vs.} community strength (Fig.~\ref{f:Spearman_diff_stats_all_life} \emph{A,C} and Fig.~\ref{f:Spearman_diff_strength_life1}), community sparsity \emph{vs.} stationarity (Fig.~\ref{f:Spearman_diff_stats_all_life} \emph{B,D}), and promiscuity \emph{vs.} force (Fig.~\ref{f:Spearman_prom}). In all cases, we show the correlation values for both the normal and tangential networks.

\end{document}